\documentclass[11pt,a4paper]{article}
\usepackage[utf8]{inputenc}
\usepackage{cite}
\usepackage{hyperref}
\usepackage{lipsum}
\usepackage{amsmath, amssymb,slashed,mathrsfs, bbm, braket}
\usepackage[pdftex,dvipsnames]{xcolor}
\usepackage{xargs}
\usepackage{tikz}
\usepackage{tikz-3dplot}
\usepackage{tikz-cd}
\usepackage{pgfplots}
\usetikzlibrary{shadows,fadings,positioning,calc,arrows,decorations.markings,patterns}

\numberwithin{equation}{section}

\newcommand{\bs}{{\mathfrak b}}

\newcommand{\cs}{{\mathfrak c}}

\newcommand{\hs}{{\mathfrak h}}

\newcommand{\qs}{{\mathfrak q}}

\newcommand\id{{\mathbbm 1}}

\begin{document}
\begin{titlepage}
  \begin{flushright} 
    UUITP -- 10/20
  \end{flushright}
  \vskip 1.5in
  \begin{center}
    {\bf\Large{       
        Cohomological Localization of $\mathcal{N}=2$ Gauge Theories with Matter}}
    \vskip
    0.5in { {Guido Festuccia, Anastasios Gorantis, Antonio Pittelli, Konstantina Polydorou and Lorenzo Ruggeri }
    } \vskip 0.5in {\small{ 
        \emph{Department of Physics and Astronomy, Uppsala University,\\ Box 516, SE-75120 Uppsala, Sweden} 
      }
    }
  \end{center}
  \vskip 0.5in
  \baselineskip 16pt

  \begin{abstract}    
    We construct a large class of gauge theories with extended supersymmetry on four-dimensional manifolds with a Killing vector field and isolated fixed points. We extend previous results limited to super Yang--Mills theory to general $\mathcal{N}=2$ gauge theories including hypermultiplets. We present a general framework encompassing equivariant Donaldson–Witten theory and Pestun's theory on $S^4$ as two particular cases. This is achieved by expressing fields in cohomological variables, whose features are dictated by supersymmetry and require a generalized notion of self-duality for two-forms and of chirality for spinors. Finally, we implement localization techniques to compute the exact partition function of the cohomological theories we built up and write the explicit result for manifolds with diverse topologies.

  \end{abstract}

  \date{}
\end{titlepage}

\tableofcontents

\newpage

\section{Introduction and summary}
The analysis of $\mathcal{N}=2$ gauge theories has led to numerous advances in our understanding of the dynamics of quantum field theories at strong coupling. Moreover, it has proven to be a fruitful arena to develop interesting connections between physics and mathematics. Two broad classes of developments in the study of $\mathcal{N}=2$ gauge theories, namely the equivariant topological twisting and Pestun's supersymmetric localization on the four-sphere, were unified in a single framework in~\cite{Festuccia:2018rew}. That analysis only applied to $\mathcal{N}=2$ super Yang--Mills theories; hence, here we consider its extension to theories that include both vector and hypermultiplets.

The first class of developments mentioned above, revolves around the topological twisting of $\mathcal{N}=2$ gauge theories. Witten showed in~\cite{Witten:1988ze} that a certain twisting of $\mathcal{N}=2$ super Yang--Mills theory results in a four dimensional topological field theory whose correlators are the Donaldson invariants~\cite{Donaldson:1990kn}. This work was later generalized by~\cite{Yamron:1988qc,Anselmi:1992tz,Anselmi:1993wm,Anselmi:1994bu,Alvarez:1993yq,Alvarez:1994ii,Labastida:1995bs,Labastida:1995zj,Hyun:1995mb}, who constructed twisted versions of $\mathcal{N}=2$ gauge theories with vector and hypermultiplets. In particular, for the purposes of our work, we are interested in the equivariant version of the topological theory of Witten, which can be defined on manifolds admitting a torus $T^{2}$ action, and was studied in~\cite{Losev:1997tp,Lossev:1997bz,Moore:1997dj,Moore:1998et}. When specifying to the omega background, its partition function was calculated in the seminal work of Nekrasov~\cite{Nekrasov:2002qd,Nekrasov:2003rj}. For more general non-compact toric manifolds $\mathcal{M}$, the partition function is a product of Nekrasov partition functions for each of the fixed points of the torus action on $\mathcal{M}$~\cite{Nekrasov:2006}, see also~\cite{Gottsche:2006tn, Gottsche:2006bm, Gasparim:2008ri}. The extension of this analysis to compact toric manifolds has been considered both recently~\cite{Bershtein:2015xfa, Bershtein:2016mxz} and in earlier  works~\cite{Bawane:2014uka, Sinamuli:2014lma, Rodriguez-Gomez:2014eza}.

The second class of developments is related to~\cite{Pestun:2007rz}, where Pestun placed an $\mathcal{N}=2$ gauge theory on a round four-sphere preserving all eight supercharges.  He was able to show that the partition function of this theory and certain supersymmetric Wilson loop observables, can be computed exactly using localization techniques. His work was later generalized to squashed four-spheres in~\cite{Hama:2012bg, Pestun:2014mja}.  These advances led to the realization that localization techniques are a powerful tool to derive exact results in supersymmetric field theories in different dimensions (for a summary of these developments see for instance the review~\cite{Pestun:2016zxk}).

The localization result for the partition function of Pestun's theory on the four-sphere is fairly simple.  It consists of an integral over a real parameter of the product of two factors, each of which can be associated by equivariant localization to one of the poles of the sphere. These are the fixed points of a $U(1)$ action generated by squaring the supercharge used for localization. One factor includes a  Nekrasov partition function for instantons, the other a partition function for anti-instantons. Hence, the final result is structurally similar to the form of the partition functions stemming from equivariant topological twisting, which only involve Nekrasov partition functions for instantons. This relation was explained in~\cite{Festuccia:2018rew} via the construction of a general class of $\mathcal{N}=2$ supersymmetric Yang--Mills theories, on any four-manifold that possesses a Killing vector field with isolated fixed points. Following~\cite{Festuccia:2018rew}, we call \emph{plus fixed points} those where instantons contribute to the partition function, while we call \emph{minus fixed points} those where anti-instantons contribute. Thus, Pestun's theory on the four-sphere and the equivariant topological twist are both examples of this general construction.  All the theories considered in~\cite{Festuccia:2018rew} can be written in terms of cohomological (twisted) fields which helps elucidating what data, of geometrical or other origin, their supersymmetric observables can depend on. These results were formulated in a more rigorous mathematical framework in~\cite{Festuccia:2019akm}.

\paragraph{Summary of results.}

In the present work we extend the results of~\cite{Festuccia:2018rew, Festuccia:2019akm} to theories involving hypermultiplets.  The inclusion of hypermultiplets generically requires the manifold to admit a spin structure. Nevertheless, this substantially enlarges the set of supersymmetric field theories which can be analyzed using localization techniques.  We reformulate the theories we construct in terms of cohomological (twisted) fields. A generalization of the notion of self-duality for two-forms was found necessary in~\cite{Festuccia:2018rew} to define the cohomological SYM theory. Similarly here we find that a generalization of a Weyl spinor is necessary to define cohomological variables for the hypermultiplets. The use of such spinors is forced upon us by supersymmetry, thus establishing an interesting relation between supersymmetry and the geometry of the manifolds where our theories live.  Furthermore, we employ localization techniques to compute the partition function of the cohomological theories we constructed.  Computations of path integrals via a purely cohomological formulation of supersymmetry appeared for  theories defined on specific manifolds in 3d~\cite{Kallen:2011ny,Assel:2016pgi,Pittelli:2018rpl}, 4d~\cite{Closset:2013sxa,Bawane:2014uka,Festuccia:2016gul,Bawane:2017gjf,Longhi:2019hdh,Festuccia:2019akm},  5d ~\cite{Kallen:2012cs,Kallen:2012va,Qiu:2013pta,Qiu:2013aga,Qiu:2014cha},  and 7d~\cite{Minahan:2015jta,Polydorou:2017jha,Iakovidis:2020znp}.   We refer to~\cite{Pestun:2016zxk} and to references therein for an exhaustive bibliography. In this paper we aim to present a general formula for the partition function of $\mathcal N=2$ gauge theories with matter on any simply connected Riemannian spin manifold admitting an isometry with isolated fixed points as well as non-trivial fluxes.  Technically, the computation of the relevant one-loop contributions descends from the index of a transversally elliptic differential operator of Dirac-type. Combining the result of this paper with those of~\cite{Festuccia:2018rew}, we find
\begin{align}\label{eq: intromasterformula}
  Z^{\mathcal N=2}_{\vec \epsilon_1 , \vec \epsilon_2 } (q , \overline q)   = \sum_{k_i \text{ discrete} } \int_{ \bf h} d a_0 \,
  &  e^{- S_{\rm cl}}   \prod_{i=1}^p Z^{\rm inst}_{\epsilon^{(i)}_1 , \epsilon^{(i)}_2} ( a_0 , k_i , q)  Z^\text{VM}_{\epsilon^{(i)}_1 , \epsilon^{(i)}_2}( a_0 , k_i ) Z^\text{HM}_{\epsilon^{(i)}_1 , \epsilon^{(i)}_2}( a_0 , k_i )   \nonumber \\
  &  \times   \prod_{i=p+1}^l Z^\text{anti-inst}_{\epsilon^{(i)}_1 , \epsilon^{(i)}_2}( a_0, k_i , \overline q ) \tilde Z^\text{VM}_{\epsilon^{(i)}_1 , \epsilon^{(i)}_2}( a_0 , k_i )  \tilde Z^\text{HM}_{\epsilon^{(i)}_1 , \epsilon^{(i)}_2}( a_0 , k_i ) ~ ,
\end{align}
where the formula above holds for a manifold with $p$ plus points and $(l-p)$ minus fixed points. In (\ref{eq: intromasterformula}), $Z^{\rm inst}$, $Z^\text{VM}$ and $Z^\text{HM}$ respectively are the instantons contribution, the vector multiplet 1-loop determinant and the hypermultiplet 1-loop determinant at a plus fixed point. Analogously,  $Z^{\rm anti-inst}$, $\tilde Z^\text{VM}$ and $\tilde Z^\text{HM}$ respectively are the anti-instantons contribution, the vector multiplet 1-loop determinant and the hypermultiplet 1-loop determinant at a minus fixed point.  The integral is taken over the Cartan gauge subalgebra ${\bf h}$, while $\overline q, q$ are counting parameters labeling (anti-)instantons. The constants $\epsilon^{(i)}_{1,2}$ are real equivariant parameters. The 1-loop contributions $ Z^\text{HM}_{\epsilon^{(+)}_1 , \epsilon^{(+)}_2}( a_0 , k_+ ) $ and $ \tilde Z^\text{HM}_{\epsilon^{(i)}_1 , \epsilon^{(i)}_2}( a_0 , k_i ) $ are Barnes double gamma functions~\cite{FRIEDMAN2004362}. With a specific choice of regularization (other choices are considered in section~\ref{sec:5}) we get for instance:
\begin{align}
  Z^\text{HM}_{\epsilon^{(+)}_1 , \epsilon^{(+)}_2}( a_0 , k_+ )    = \prod_{\rho \in \mathcal R} \Gamma_2 ( i \, \rho(\Phi_0) + ( (\epsilon^{(+)}_1 + \epsilon^{(+)}_2)/2 )  |  \epsilon^{(+)}_1 , \epsilon^{(+)}_2 )  ~ ,
\end{align}
where we have a product over hypermultiplet representations $\mathcal R$, while the fugacity $\Phi_0 $ at a plus point is a combination of the Coulomb branch parameter $a_0$ and the function $k_+$ encoding the flux contribution:
\begin{equation}
  \Phi_0 =   a_0  +  k_+( \epsilon_1^{(+)}  ,  \epsilon_2^{(+)}  )  ~ .
\end{equation}
As we spell out in the main text,  the expression for $\tilde  Z^\text{HM}_{\epsilon^{(-)}_1 , \epsilon^{(-)}_2}( a_0 , k_- ) $ and for $\Phi_0'$ valid at a minus fixed point are analogous.  Hypermultiplet masses enter the partition function as a constant shift of $a_0$. At the level of the Lagrangian,  a massless  hypermultiplet is made massive by weakly gauging   a   $U(1)$ flavour symmetry, where  the mass coincides with the real part of a  scalar from a background vector multiplet.

\paragraph{Outline of the paper.}

In section~\ref{sec:two}, we review how to place an $\mathcal{N}=2$ field theory on Euclidean four-manifolds preserving some supersymmetry. This can be accomplished by coupling the theory to a supersymmetric rigid supergravity background, whose properties we review. In particular, we show that the usual hypermultiplet action, quadratic in derivatives, when coupled to rigid supergravity, is $\delta$-exact up to total derivatives.

In section~\ref{secspin}, we consider a Riemannian spin four-manifold $\mathcal{M}$ that admits a Killing vector field with isolated fixed points. We review the arguments in~\cite{Festuccia:2018rew} showing that $\mathcal{M}$ admits globally well defined Killing spinors that depend on a choice of a plus/minus label for each fixed point. In order to deal with off-shell hypermultiplets, we also need to show the existence of auxiliary Killing spinors satisfying certain properties. We prove that a smooth, globally well-defined choice for these auxiliary Killing spinors exists.

In section~\ref{sec:4}, we move to reformulate the gauged hypermultiplet in terms of cohomological fields. For this we first define novel splits of the Dirac spinor bundle into two subbundles.  These splits require the existence of a vector field with isolated fixed points and depend on the choice of $\pm$ at each fixed point. We also study the relation between these spinor bundles and the ``flipping'' sub-bundles of the bundle of two-forms studied in~\cite{Festuccia:2018rew}. The appropriate cohomological fields for the hypermultiplet are elements of the novel spinor bundles we introduced. We explicitly prove that there is a smooth invertible map between cohomological variables and the usual hypermultiplet component fields. We also show how supersymmetry organizes the cohomological fields in different multiplets. Finally we rewrite the action for a gauged hypermultiplet in terms of the cohomological fields. In the reformulation of the theory in terms of cohomological variables, there is an important difference with~\cite{Festuccia:2018rew}, where only vector multiplets were considered. In the case of the hypermultiplet the twisted fields are spinors so that generically the four-manifold has to be spin. This is however too restrictive. For instance, depending on the flavor symmetry of the theory, it may be enough for the manifold to be spin$^c$.

Finally in section~\ref{sec:5}, having constructed the cohomological theory, we use it to set up the localization computation of the partition function.  As an application, we apply our formula to the specific case of the squashed four-sphere. We find perfect agreement with the localization result for $\mathcal{N}=2$ matter multiplets found in ~\cite{Hama:2012bg}.

\paragraph{Outlook.}

A compelling direction to be explored in a future work would be applying the technology developed in this paper to study the consequences of S-duality invariance of $\mathcal N=4$ super Yang--Mills theory. Indeed, it should be possible to generalize the analysis of~\cite{Vafa:1994tf}, valid for pure topological twisting, to the case of equivariant topological twisting by rearranging the twisted fields of vector and hypermultiplets in different cohomological complexes.

Another intriguing line of investigation   would be deriving  the cohomological complex for $\mathcal N=4$ theories in three dimensions. Although this can simply be achieved by means of dimensional reduction, the outcome would be non-trivial as several fields of the three-dimensional cohomological complex would become charged under an $SU(2)$ (Coulomb) R-symmetry. This would open up the possibility of more elaborate topological twisting, as well as of exploring three-dimensional mirror symmetry from a cohomological viewpoint.

\section{$\mathcal{N}=2$ theories on four-manifolds}
\label{sec:two}
In this section we review the construction of rigid $\mathcal{N}=2$ supersymmetric fields theories on a curved four-manifold by coupling to background supergravity. Assuming the existence of appropriate Killing spinors, we write down the supersymmetry variations for fields in vector multiplets and hypermultiplets.  We also present a $\delta$-exact Lagrangian for a gauged hypermultiplet.

\subsection{Review of $\mathcal N=2$ rigid supergravity}
\label{sec:sugra-background}
In order to couple  a supersymmetric field theory defined in flat space to off-shell supergravity, we have to set the fermionic fields in the supergravity multiplet to zero and freeze the bosonic supergravity fields to fixed values. If this background is invariant under some supergravity variation, the resulting theory is supersymmetric~\cite{Festuccia:2011ws}.  Here we consider $\mathcal{N}=2$ theories with a conserved $SU(2)$ R-current. Their supercurrent multiplet was studied by Sohnius~\cite{Sohnius:1978pk} and the $\mathcal{N}=2$ Poincar\'e supergravity to which they couple is described in~\cite{deWit:1979dzm,deWit:1980lyi,deWit:1984rvr,Freedman:2012zz}. Rigid $\mathcal{N}=2$ supergravity backgrounds and the conditions they have to satisfy to preserve supersymmetry have been considered in~\cite{Klare:2013dka,Pestun:2014mja,Butter:2015tra}.

In order to specify the supergravity background we need the following data:
\begin{itemize}
\item A Riemannian manifold $\mathcal{M}$ equipped with a metric $g$ and a spin structure\footnote{We are going to comment on non-spin manifolds in Section \ref{sec:hypermultiplet}.}.
\item An $SU(2)_R$ connection ${{V_{\mu}}^i}_j$. (Here and in the following $i,j\ldots$ are $SU(2)_R$ indices.)
\item Various other auxiliary fields: a one-form $G_\mu$, a two-form $W_{\mu\nu}$, a scalar $N$, a closed two-form $F_{\mu\nu}$, and a scalar $S_{ij}$ transforming as an $SU(2)_R$ triplet.
\end{itemize}
The supergravity variations are parametrized by a left-handed spinor $\zeta^{i}_{\alpha}$ and a right-handed spinor $\bar{\chi}_{i}^{\dot{\alpha}}$, both transforming in the fundamental representation of the $SU(2)_{R}$ R-symmetry. Here $i$ is the $SU(2)_{R}$ index and $\alpha,\dot{\alpha}$ are spinor indices. For a brief review of the conventions we use, see Appendix~\ref{app:conv}~.  Unless otherwise noted, we take these spinors to obey symplectic-Majorana reality conditions:
\begin{equation}
  (\zeta_{i\alpha})^{*} = \zeta^{i\alpha}~, \qquad
  (\bar{\chi}_{i}^{\dot{\alpha}})^{*} = \bar{\chi}_{\dot{\alpha}}^{i}~.
  \label{eq:reality-killing}
\end{equation}
Requiring the background to be invariant under the supergravity variation parametrized by $\zeta^{i}_{\alpha}$ and $\bar{\chi}_{i}^{\dot{\alpha}}$, one obtains two sets of Killing spinor equations. The first set is
\begin{equation}
  \begin{split}
    \label{eq:killinone}
    &(D_\mu -i G_\mu) \zeta_{i}-{i\over 2} W^+_{\mu\rho} \sigma^{\rho}{\bar \chi}_{i} -{i\over 2} \sigma_\mu {\bar \eta}_i=0~, \\
    &(D_\mu + i G_\mu) {\bar \chi}^{i}+ {i\over 2} W^-_{\mu\rho} {\bar \sigma}^\rho{ \zeta^{i}} -{i\over 2} {\bar \sigma_\mu} {\eta}^i=0~,
  \end{split}
\end{equation}
where $D_{\mu}$ is a covariant derivative that incorporates the $SU(2)_{R}$ connection $V_{\mu}{}^{i}{}_{j}$. The second set is
\begin{equation}
  \begin{split}
    \label{eq:killintwo}
    &\Big(N-{1\over 6} R\Big)\bar \chi^i=
    4i \partial_\mu G_\nu \bar \sigma^{\mu\nu} \bar \chi^i
    +{i} \big(\nabla^\mu+2 i G^\mu\big) W^{-}_{\mu\nu} \bar \sigma^\nu\zeta^i
    +i \bar \sigma^\mu\big(D_\mu+{i} G_\mu\big) \eta^i ~,\\
    &\Big(N-{1\over 6} R\Big)\zeta_i
    =-4i \partial_\mu G_\nu \bar \sigma^{\mu\nu} \zeta_i
    -{i} \big(\nabla^\mu
    -2i G^\mu\big) W^{+}_{\mu\nu} \sigma^\nu\bar \chi_i
    +i \sigma^\mu\big(D_\mu-{i} G_\mu\big) \bar \eta_i ~,
  \end{split}
\end{equation}
where $R$ is the Ricci scalar and the spinors $\eta^{i}$ and $\bar{\eta}^{i}$ are defined as:
\begin{equation}
  \begin{split}
    & \eta_i=({\mathcal F}^+-W^+) \zeta_i-2 G_\mu\sigma^\mu \bar \chi_i- S_{i j} \zeta^j~, \\[2pt]
    & \bar \eta^i= -({\mathcal F}^-- W^-) \bar \chi^i +2 G_\mu \bar \sigma^\mu \zeta^i-S^{i j} \bar \chi_j~.
  \end{split}
\end{equation}
Here we use the notation $W^+={1\over 2} W_{\mu\nu} \sigma^{\mu\nu}$ and $W^-={1\over 2} W_{\mu\nu}\bar \sigma^{\mu\nu}$ (similarly for $\mathcal F$).

The spinors $\zeta^{i}$ and $\bar \chi_{i}$ can be used to construct various bilinears. Restricting our attention to singlets of the $SU(2)$ R-symmetry, we have the scalars
\begin{equation}
  \label{stsdef}
  s = 2 \zeta^i \zeta_i~,\qquad
  \tilde{s} = 2\bar \chi^i \bar \chi_i~, 
\end{equation}
and the vector field
\begin{equation}    
  v^{\mu} = 2\bar \chi^i\bar\sigma^\mu \zeta_i~. \label{eq:Killing-vector}
\end{equation}

The reality conditions~\eqref{eq:reality-killing} imply that $v^\mu,~s$ and $\tilde s$ are real and that $s$ and $\tilde s$ are nowhere negative. The vector $v$ and the scalars $s,~\tilde s$ satisfy $||v||^2=s\tilde s$, hence they are not independent.  Using the Killing spinor equations we can show that $v^\mu$ is a Killing vector and that $s,~\tilde s$ are constant along the orbits of $v$. We assume that both $\sqrt s$ and $\sqrt{\tilde s}$ are smooth in a neighborhood of the fixed points.

\subsection{Supersymmetric multiplets}
\label{sec:multiplets}
Here we assume that the Killing spinor equations introduced in the previous subsection are satisfied in some supergravity background. We present the structure of the supersymmetry variations for vector multiplets and hypermultiplets coupled to this background.

\subsubsection{Vector multiplet}
The $\mathcal{N}=2$ vector multiplet contains a complex scalar field $X$, a gauge field $A_{\mu}$, two gauginos $\lambda_{i\alpha}$ and $\tilde{\lambda}_{\dot{\alpha}}^{i}$ that transform in the fundamental of $SU(2)_{R}$ and an auxiliary scalar field $D_{ij}$ transforming as a triplet of $SU(2)_{R}$. All these fields (except $A_{\mu}$) transform in the adjoint representation of the gauge group.  The supersymmetry variations are given by:
\begin{eqnarray}
  \label{vmulvr}
  &&\hspace{-15pt}
     \delta {\bar X} = {\bar \chi}^i{\bar \lambda_i}~,
     \qquad
     \delta { X} = -{ \zeta}_i{\lambda^i} ~, \nonumber \\
  &&\hspace{-15pt}
     \delta A_{\mu} = i\zeta_i \sigma_\mu {\bar \lambda}^i
     +i {\bar \chi}^i {\bar \sigma}_\mu \lambda_i~, \nonumber \\
  &&\hspace{-15pt}
     \delta D_{i j} = i \zeta_i \sigma^\mu \big( D_\mu\! +i  G_\mu \big){\bar \lambda}_j
     -i {\bar \chi}_i \bar \sigma^{\mu}\big( D_\mu\! -i G_\mu \big) \lambda_j
     +2i  [X,\bar \chi_i \bar \lambda_j]
     +2i  [\bar X, \zeta_i \lambda_j]
     + (i \leftrightarrow j)~, \\
  &&\hspace{-15pt}
     \delta \lambda_{i} = -2 i (D_\mu-2iG_\mu)  X {\sigma^\mu{\bar \chi}_i}\!
     +2\big(F^+\!-{\bar X}\, W^+\big) \zeta_i+D_{i j}\zeta^j
     +2 i  [X,\bar X] \zeta_i-2 X\eta_i~,\nonumber \\
  &&\hspace{-15pt}
     \delta {\bar \lambda}^{i} = 2 i (D_\mu +2i G_\mu){\bar X} {\bar \sigma^\mu\zeta^i}\!
     +\!2 \big(F^-\!- X\, W^-\big)\bar \chi^i-D^{i j}\bar \chi_j
     -2i  [X, \bar X] \bar \chi^i +2 \bar X \bar \eta^i~. \nonumber
\end{eqnarray}
We used the shorthand notation $F^+={1\over 2 } F_{\mu\nu} \sigma^{\mu\nu}$ and $F^-={1\over 2 } F_{\mu\nu} \bar \sigma^{\mu\nu}$ where $F_{\mu\nu}$ is the field strength for the gauge field $A_\mu$.

The square supersymmetry variation of a field $\Psi$ in the vector multiplet results in a translation along the vector field $v$ defined in~\eqref{eq:Killing-vector}, together with a gauge transformation and an $SU(2)_R$ transformation
\begin{equation}
  \label{salg}
  \delta^2 \Psi=   i {\cal L}_{v} \Psi+ i v^\mu V_\mu \circ \Psi
  +i \Lambda^{\!(R)} \!\!\circ \Psi-i [\Phi,\Psi]~.
\end{equation}
Here ${\cal L}_{v}$ is the Lie derivative along $v$, and $\circ$ denotes that $\Psi$ is acted upon according to which $SU(2)_R$ representation it belongs. The gauge transformation parameter is
\begin{equation}
  \label{eq:Phidef}
  \Phi = i v^\mu A_\mu+s \bar X+\tilde s X~,
\end{equation} 
and $\Lambda^{\!(R)}$ is a $SU(2)_R$ transformation parameter given by:
\begin{equation}
  \label{eq:Aambda}
  \Lambda^{\!(R)}_{i j} = \bar \chi_i \bar \sigma^\mu (D_\mu -i G_\mu)\zeta_j
  -\zeta_i \sigma^\mu (D_\mu+iG_\mu) \bar \chi_j+ (i\leftrightarrow j)~.
\end{equation}

\subsubsection{Hypermultiplet}
For the following we will embed the gauge group in $Sp(k)$ and consider a hypermultiplet in the fundamental of this $Sp(k)$. The hypermultiplet contains a scalar $q_{ni}$ (where the index $n$ transforms under $Sp(k)$ and runs over $n=1,\ldots,2k$, while the index $i$ transforms under the fundamental of $SU(2)_{R}$), and a pair of spinors $\psi_{\alpha n}$ and $\bar{\psi}_{{\dot \alpha} n}$. Additionally there are auxiliary fields $F_{n\check \imath}$ that are necessary for the off-shell closure of the supersymmetry algebra. The $F_{n\check \imath}$ transform in the fundamental of an $SU(2)_{\check R}$ symmetry that is generically distinct from the $SU(2)$ R-symmetry.

We take the Grassmann-even fields in the hypermultiplet to satisfy the following reality conditions:
\begin{equation}
  \label{realhyp}
  (q_{n i})^*= q^{n i}~, \qquad (F_{n \check \imath})^*= F^{n \check \imath}~.
\end{equation}

The supersymmetry variations of the hypermultiplet components are:
\begin{equation}
  \begin{split}
    \delta q_{ni}
    &= \zeta_{i} \psi_{n} + \bar{\chi}_{i}\bar{\psi}_{n}~, \\
    \delta\psi_{n}
    &= 2i(D_{\mu}q_{ni})\sigma^{\mu}\bar{\chi}_{i}
    +iq_{ni}\sigma^{\mu}
    \left(D_{\mu}+ i G_{\mu}\right)\bar{\chi}_{i}
    +4i\bar{X}_{n}{}^{m}q_{mi}\zeta^{i}
    +2iF_{n\check \imath}\check{\zeta}^{\check \imath}~,\\
    \delta\bar{\psi}^{n}
    &= 2i(D_{\mu}q^{ni})\bar{\sigma}^{\mu}\zeta_{i}
    +iq^{ni}\bar{\sigma}^{\mu}
    \left(D_{\mu} - iG_{\mu}\right)\zeta_{i}
    +4iX^{n}{}_{m}q^{mi}\bar{\chi}_{i}
    +2iF^{n\check \imath}\check{\bar{\chi}}_{\check \imath}~,\\
    \delta F_{n\check \imath}
    &= \check{\zeta}_{\check \imath}\left[
      \sigma^{\mu}\left(D_{\mu} - iG_{\mu}\right)\bar{\psi}_{n}
      -2X_{n}{}^{m}\psi_{m}
      +2(\lambda^{j})_{n}{}^{m}q_{mj}
      - iW^{+}\psi_{n}
    \right] \\
    &\qquad+\check{\bar{\chi}}_{\check \imath}\left[
      \bar{\sigma}^{\mu}\left(D_{\mu} + iG_{\mu}\right)\psi_{n}
      +2\bar{X}_{n}{}^{m}\bar{\psi}_{m}
      -2(\bar{\lambda}^{j})_{n}{}^{m}q_{mj}
      + iW^{-}\bar{\psi}_{n}
    \right]~,
  \end{split}
\end{equation}
where $X_{n}{}^{m}=X^{\alpha}{{t^{\alpha}}_n}^m$ (and similarly for other vector multiplet components) and the derivative $D_{\mu}$ is covariant with respect both to the gauge symmetry and the R-symmetry. Finally, we introduced the checked spinors $\check{\zeta}_{\check \imath}$ and $\check{\bar{\chi}}_{\check \imath}$, which need to satisfy the constraints:
\begin{equation}
  \begin{aligned}
    & \zeta_{i}\check{\zeta}_{\check \jmath} - \bar{\chi}_{i} \check{\bar{\chi}}_{\check \jmath} = 0 ~ ,
    &\qquad
    & \check{\zeta}_{\check \imath} \check{\zeta}^{\check \imath}= \bar \chi^i \bar \chi_i  ~ , \\
    & \check{\bar{\chi}}^{\check \imath} \bar{\sigma}^{\mu} \check{\zeta}_{\check \imath} +\bar \chi^i\bar\sigma^\mu \zeta_i =0~ ,
    &
    & \check{\bar{\chi}}_{\check \imath}\check{\bar{\chi}}^{\check \imath} = \zeta^i\zeta_i ~ ,
  \end{aligned}
\end{equation}
for the off-shell closure of the supersymmetry algebra. Unless otherwise noted, we assume that $\check \zeta_{\check \imath}$ and $\check {\bar \chi}_{\check \imath}$ satisfy symplectic-Majorana reality conditions:
\begin{equation}
  (\check{\zeta}^{\check \imath}_{\alpha})^{*} = \check{\zeta}_{\check \imath}^{\alpha}~,\qquad
  (\check{\bar{\chi}}^{\check \imath \dot{\alpha}}) ^{*} =  \check{\bar{\chi}}_{\check \imath\dot{\alpha}}~.
  \label{eq:reality-killing}
\end{equation}

Except when acting on the auxiliary fields $F_{n\check \imath}$, the square supersymmetry variation of a field $\Psi$ of the hypermultiplet, is given by:
\begin{equation}
  \label{eq:square-susy-original}
  \delta^{2} \Psi =  
  i {\mathcal L}_{v} \Psi+ i v^\mu V_\mu \circ \Psi +i \Lambda^{\!(R)} \!\!\circ \Psi + \mathcal{G}_{\Phi}\diamond\Psi~.
\end{equation}
This includes a translation along $v^{\mu}=2\bar \chi^i\bar\sigma^\mu \zeta_i$, an $SU(2)_R$ transformation and a gauge transformation. The gauge transformation parameter $\Phi$ and the $SU(2)_R$ transformation parameter $\Lambda^{\!(R)}$ are as in~\eqref{eq:Phidef} and~\eqref{eq:Aambda}.  The squared supersymmetry variation of the auxiliary fields similarly includes $SU(2)_{\check R}$ transformations:
\begin{equation}
  \label{eq:square-susy-aux}
  \delta^{2} \Psi =  
  i {\mathcal L}_{v} \Psi+ i v^\mu {\check V}_\mu \circ \Psi  + i \Lambda^{\!(\check R)} \!\!\circ \Psi + \mathcal{G}_{\Phi}\diamond\Psi~.
\end{equation}
Here ${{{~\check V_\mu}}}$ is a background connection for $SU(2)_{\check R}$ and the $SU(2)_{\check R}$ transformation parameter $\Lambda^{\!(\check R)} $ is given by:
\begin{align}
  \label{sqau}
  \Lambda^{(\check R)}_{\check \imath \check \jmath}= 2 {\check \zeta_{\check \imath}}{ \sigma}^\mu \Big({\check D}_\mu- {i} G_\mu\Big) {\check{\bar \chi}_{\check \jmath}}+2 i {\check \zeta_{\check \imath}}W^+{\check \zeta_{\check \jmath}}
  -2 {\check{\bar \chi}_{\check \imath}}{\bar \sigma}^\mu \Big({\check D}_\mu+ {i} G_\mu\Big) {\check \zeta}_{\check \jmath} +2i{\check{\bar \chi}_{\check \imath}}W^-{\check{\bar \chi}_{\check \jmath}}+\big({\check \imath \leftrightarrow \check \jmath}\big)~.
\end{align}
The derivative ${\check D}_\mu$ is covariant with respect to the background $SU(2)_{\check R}$ connection ${{{~\check V_\mu}}}$~. Note that in~\eqref{eq:square-susy-aux}, the connection ${{{~\check V_\mu}}}$ cancels between the terms $ i v^\mu {\check V}_\mu \circ \Psi $ and $ i \Lambda^{\!(\check R)} \!\!\circ \Psi$~.

\subsection{Hypermultiplet Lagrangian}
\label{sec:gaug-hyperm}
Having completed the coupling to rigid supergravity, we can write a supersymmetric Lagrangian $ \mathcal{L} = \mathcal{L}_{B} + \mathcal{L}_{F}$ for the hypermultiplet:
\begin{subequations}\label{eq:original_action}
  \begin{align}
    \mathcal{L}_{B}
    &=  + \frac{1}{2}(D^{\mu}q^{ni})(D_{\mu}q_{ni})
      -  \frac{i}{2} q^{n}{}_{i}(D^{ij})_{n}{}^{m}q_{mj}
      + \frac{1}{2}F^{n\check{\imath}}F_{n\check{\imath}} \nonumber\\
    &\qquad
      - \left(\frac{R}{12} + \frac{N}{4}\right)q^{ni}q_{ni}
      + q^{ni}\{\bar{X},X\}_{n}{}^{m}q_{mi}~, \\
    \mathcal{L}_{F}
    &= -  \frac{i}{2}\psi^{n}\sigma^{m}\left(
      D_{\mu} - i G_{\mu}\right)\bar{\psi}_{n}
      + \frac{i}{2}\psi^{n}X_{n}{}^{m}\psi_{m}
      + \frac{i}{2}\bar{\psi}^{n}\bar{X}_{n}{}^{m}\bar{\psi}_{m} \nonumber\\
    &\qquad
      - i\psi^{n}(\lambda^{i})_{n}{}^{m}q_{mi}
      - i\bar{\psi}^{n}(\bar{\lambda}^{i})_{n}{}^{m}q_{mi}
      - \frac{1}{4}\left(\psi^{n}W^{+}\psi_{n}
      +\bar{\psi}^{n}W^{-}\bar{\psi}_{n}\right)~.
  \end{align}
\end{subequations}
It turns out that (up to total derivatives) this Lagrangian is itself the supersymmetry variation of some Grassmann-odd $V_{G}$:
\begin{equation}
  \mathcal{L} = \mathcal{L}_{B} + \mathcal{L}_{F} = \delta V_{G} ~,
\end{equation}
where $V_{G}$ is given by the following expression:
\begin{align}\label{eq:qexact_lagr}
  V_{G} =
  \frac{1}{2(s+\tilde{s})}
  &\Big[
    2 i (D_\mu + i G_\mu)(q_{ni}\zeta^i)\sigma^\mu \bar{\psi}^n
    - 2i(D_\mu - i G_\mu)(q_{ni}\bar{\chi}^i)\bar{\sigma}^\mu \psi^n
    + 2 i F_{n\check\imath}(\check{\bar{\chi}}^{\check\imath} \bar{\psi}^n -\check{\zeta}^{\check\imath} \psi^n)\nonumber\\
  &-4i q_{mi} (X^m {}_n \bar{\chi}^i \bar{\psi}^n + \bar{X}^m {}_n \zeta^i \psi^n)
    - 2 q_{ni}(\bar{\chi}^i W^- \bar{\psi}^n + \zeta^i W^+ \psi^n)\nonumber\\
  &- \frac{2}{s+\tilde{s}}v^{\nu}\mathcal{F}_{\mu\nu}
    q_{ni}(\bar{\chi}^{i}\bar{\sigma}^{\mu}\psi^{n} - \zeta^{i}\sigma^{\mu}\bar{\psi}^{n})
    - 4i q^{ni} \left[(\lambda_{i})_{n}{}^{m}\zeta_{j}
    + (\bar{\lambda}_{i})_{n}{}^{m}\bar{\chi}_{j} \right]
    q_{m}{}^{j}
    \Big]~.
\end{align}

The expression for $V_{G}$ in~\eqref{eq:qexact_lagr} is in agreement with and generalizes a similar one found in~\cite{Hama:2012bg}, to which it reduces if we assume that $s+\tilde s$ is a constant (in that case $\mathcal{F}_{\mu\nu}=0$).

\section{Construction of Killing spinors}
\label{secspin} 
In the last section we have considered gauged hypermultiplets coupled to a rigid $\mathcal{N}=2$ supergravity background. We assumed that this background preserves supersymmetry, i.e. that there are nonzero solutions to the Killing spinor equations~\eqref{eq:killinone} and~\eqref{eq:killintwo}.  A general class of backgrounds allowing solutions to these equations was studied in~\cite{Festuccia:2018rew}. Here we will briefly review their main properties.

Consider a Euclidean orientable four-manifold $\mathcal{M}$ with metric $g$ and a spin structure. It was shown in~\cite{Festuccia:2018rew} that the Killing equations can be satisfied by spinors $\zeta^{i}$ and $\bar \chi_{i}$ that are both non-vanishing provided that the metric admits a Killing vector $v$ whose fixed points are isolated.  In addition to $v$, the supergravity background is specified by a choice of a real, nowhere-negative scalar $s$ on $\mathcal{M}$ that is constant along orbits of $v$. Moreover $s$ has to approach a positive constant $K$ at a subset of the fixed points of $v$ and needs to go to zero as $||v||^2/K$ at the remaining fixed points.  At these fixed points of $v$, the scalar $\tilde s=||v||^2 /s$ approaches $K$. Hence,  the fixed points of $v$ are separated into a set where $\tilde s=0$ with $s=K$ and a second set where $s=0$ and $\tilde s=K$. The fixed points satisfying $\tilde{s}=0$ are plus fixed points, whereas those satisfying $s=0$ are minus fixed points. This is consistent with~\cite{Festuccia:2018rew}.

\subsection{Killing spinors}
\label{sec:constr-killing-spinors}
In this subsection, we employ the Killing vector $v$ and the scalar $s$ to construct spinors $\zeta^i_\alpha$ and $\bar \chi^i_{\dot \alpha}$ that satisfy the reality conditions $({\zeta}_{i \alpha})^*= {\zeta}^{i \alpha}$ and $(\bar \chi_i^{\dot \alpha})^*=\bar \chi^i_{\dot\alpha} $ and such that
\begin{equation}
  \label{bilspin}
  \zeta^i \zeta_i={s\over 2}~,\qquad
  \bar \chi^i \bar \chi_i={\tilde s\over 2}~, \qquad
  \bar \chi^i\bar\sigma^\mu \zeta_i ={1\over 2} v^\mu~.
\end{equation}

We cover the manifold with charts $U_k$, such that every fixed point of $v$ belongs to a single distinct chart and we make a choice of vielbein $e^{a}{}_{k}$ in each chart. We can also assume that there are no overlaps between charts containing different fixed points of $v$. We consider the spinors $\zeta^{i}$ and $\bar \chi_{i}$ given by the following expressions in every chart:
\begin{equation}
  \label{defchi}
  \zeta^i_\alpha={\sqrt{s}\over 2}\, \delta^i_\alpha~,\qquad
  {\bar \chi}_i=  {1\over s} v^\mu \bar \sigma_\mu  \zeta_i~.
\end{equation}
These spinors satisfy the reality conditions and the constraints~\eqref{bilspin}. 

In going from chart to chart, $\zeta$ transforms under $SU(2)_l\times_{\mathbb{Z}_2} SU(2)_R$. For the form of $\zeta$ above to be valid in each chart, we have to undo the $SU(2)_l$ transformation by an appropriate $SU(2)_R$ transformation. The expression for $\bar\chi$ will then also be valid in each chart because it is directly related to $\zeta$.

Unfortunately~\eqref{defchi} is singular in the charts containing a fixed point of $v$ where $s=0$. To fix this problem, in going from a chart where $s\neq 0$ everywhere to a chart where there is a fixed point of $v$ with $s=0$, we can act with a further $SU(2)_R$ transformation:
\begin{equation}
  \label{suttr}
  {U_i}^j= i {v^\mu \over ||v||}{{\sigma_\mu}_i}^j~.
\end{equation}
As a consequence, in charts containing a fixed point where $s=0$, the spinors are:
\begin{equation}
  \label{newspin}
  {\bar \chi}_i^{\dot \alpha}= -i {\sqrt{\tilde s}\over 2}\delta^{\dot \alpha}_i~,\qquad \zeta_i= -{1\over \tilde s}v^\mu \sigma_\mu \hat{\bar \chi}_i~.
\end{equation}

This specifies regular spinors $\zeta$ and $\bar \chi$ on ${\cal M}$ that satisfy the reality conditions and for which the relations~\eqref{bilspin} are satisfied. Moreover there is a choice of smooth background supergravity fields for which the spinors $\zeta$ and $\chi$, that we just constructed, satisfy the Killing spinor equations~\eqref{eq:killinone}\eqref{eq:killintwo}~(see~\cite{Festuccia:2018rew}). The resulting expressions for the background fields are presented in Appendix~\ref{sec:sugra-background-solutions}.

\subsection{Auxiliary Killing spinors}
An additional element we need to consider is the construction of smooth auxiliary spinors $\check{\zeta}^{\check{\imath}}_{\alpha}$ and $\check{\bar{\chi}}_{\check{\imath}}^{\dot{\alpha}}$, since they are used for the off-shell extension of the supersymmetry transformations for the $\mathcal{N}=2$ hypermultiplet. These spinors satisfy the reality conditions:
\begin{equation}
  (\check{\zeta}_{\check{\imath}}^\alpha)^{*} = \check{\zeta}^{\check{\imath}}_\alpha~,\qquad
  (\check{\bar{\chi}}^{\check{\imath} \dot{\alpha}}) ^{*} = \check{\bar{\chi}}_{\check{\imath}\dot{\alpha}}~,
\end{equation}
as well as the following constraints:
\begin{equation}
  \begin{split}
    \label{eq:1}
    &\zeta_{i}\check{\zeta}_{\check{\jmath}} - \bar{\chi}_{i} \check{\bar{\chi}}_{\check{\jmath}} = 0 ~ , \qquad\qquad
    \check{\zeta}_{\check{\imath}} \check{\zeta}^{\check{\imath}}= \frac{\tilde{s}}{2}  ~ , \\
    &\check{\bar{\chi}}^{\check{\imath}} \bar{\sigma}^{\mu} \check{\zeta}_{\check{\imath}}
    = - \frac{1}{2} v^{\mu} ~ , \qquad\qquad
    \check{\bar{\chi}}_{\check{\imath}}\check{\bar{\chi}}^{\check{\imath}} = \frac{s}{2} ~ .
  \end{split}
\end{equation}
These spinors transform under $SU(2)_{l} \times_{\mathbb{Z}_{2}} SU(2)_{\check{R}}$ and $SU(2)_{r} \times_{\mathbb{Z}_{2}} SU(2)_{\check{R}}$ respectively. The $SU(2)_{\check{R}}$ bundle is not generally identified with the $SU(2)_{R}$ bundle. The constraints in~\eqref{eq:1} determine $\check\zeta$ and $\check{\bar \chi}$ uniquely up to local $SU(2)_{\check{R}}$ transformations.

The construction of these spinors parallels that of $\zeta$ and $\chi$. On patches $U_{k}$ that do not contain fixed points where $s=0$, we define the spinors as follows:
\begin{equation}
  \label{conchecks}
  {\check{\bar{\chi}}}_{\check{\imath}}^{\dot{\alpha}}
  = \frac{\sqrt{s}}{2}\delta^{\dot{\alpha}}_{\check{\imath}}~,
  \qquad
  \check{\zeta}_{\check{\imath}\, \alpha}
  = -\frac{1}{s} v^{\mu}\left(\sigma_{\mu}\check{\bar{\chi}}_{\check{\imath}}\right)_{\alpha}~.
\end{equation}
It can be checked using the expressions for $\zeta^{i}$ and $\bar\chi_{i}$ in~\eqref{defchi}, that these satisfy the constraints~\eqref{eq:1}.  In this case in going from patch to patch, we have to undo $SU(2)_r$ transformations by appropriate $SU(2)_{\check R}$ transformations.

On a patch $U_{l}$ that includes a fixed point of $v$ with $s=0$, we take instead:
\begin{equation}
  \check{\zeta}_{\alpha}^{\check{\imath}}
  = i \frac{\sqrt{\tilde{s}}}{2}\delta_{\alpha}^{\check{\imath}}~,
  \qquad
  \check{\bar{\chi}}_{\check{\imath}}^{\dot{\alpha}}
  = \frac{1}{\tilde{s}} v^{\mu}\left(\bar{\sigma}_{\mu}\check{\zeta}_{\check{\imath}}\right)^{\dot{\alpha}}~,
\end{equation}
which also satisfy the constraints~\eqref{eq:1} with $\zeta^{i}$ and $\bar \chi_{i}$ in~\eqref{newspin}.  As before, when transitioning to one of these patches, there is an extra $SU(2)_{\check R}$ transformation given by:
\begin{equation}
  U_{\check{\imath}}{}^{\check{\jmath}} = i \frac{v^{\mu}}{\vert\vert v \vert\vert } (\sigma_{\mu})_{\check{\imath}}{}^{\check{\jmath}}~.
\end{equation}
Hence, we have found smooth solutions to the constraints~\eqref{eq:1}. There is a different construction of solutions to these constraints that is often used in the literature. This is given by:
\begin{equation}
  \check{\zeta}_{\alpha}^{\check{\imath}}
  = i \sqrt{\tilde{s}\over s}\,\delta^{\check{\imath}}_i \,\zeta^i_\alpha~,
  \qquad
  \check{\bar{\chi}}_{\check{\imath}}^{\dot{\alpha}}
  = i  \sqrt{s\over \tilde s}\,\delta_{\check{\imath}}^i \,{\bar \chi}_i^{\dot \alpha}~.
\end{equation}
In this case $SU(2)_{\check R}$ is identified with $SU(2)_R$; however the resulting spinors are singular at the fixed points of $v$.

\section{Twisted supersymmetry}
\label{sec:4}
In this section we rewrite the component fields for the vector multiplet and hypermultiplet in terms of twisted variables. The case of the vector multiplet was considered in~\cite{Festuccia:2018rew}. The construction of these twisted variables is intimately connected with the geometry of ${\mathcal M}$. In particular, for the vector multiplet it relies on a novel decomposition of the bundle of two-forms on ${\mathcal M}$ which we review. For the hypermultiplet we will introduce a corresponding decomposition of spinors on ${\mathcal M}$.

\subsection{Flipping projectors}
Consider the bundle of two-forms $\Lambda^2({\mathcal M})$. On an orientable Euclidean manifold with metric $g$, we can split $\Lambda^2({\mathcal M})$ in two orthogonal sub-bundles, consisting of self-dual and anti self-dual two-forms. Self-dual two-forms play an essential role in Donaldson--Witten theory and its equivariant extension~\cite{Witten:1988ze, Losev:1997tp}. As shown in~\cite{Festuccia:2018rew}, on a manifold equipped with a vector field $v$ with isolated fixed points, one can introduce a different splitting of the bundle of two-forms. This splitting arises naturally in the study of the ${\mathcal N}=2$ theories on ${\mathcal M}$ we presented in section~\ref{sec:two}.

Let $v$ be a real vector field on ${\mathcal M}$ (with isolated fixed points) and for what follows we will take it to be the Killing vector. The $s$ and $\tilde s$ , which are the spinor bilinears in~\eqref{stsdef}, determine the $\pm$ fixed points. We can define projectors $P_\pm$ on the bundle of two-forms as follows:
\begin{align}\label{defpp}
  & P_+={1\over 2(s^2+\tilde s^2)}\left((s+\tilde s)^2 \id
    +(s^2-\tilde s^2) \star -4 \kappa\wedge \iota_v\right) ~, & P_- = \id - P_+ ~ ,
\end{align}
where we use $\id$ to indicate the identity operator throughout the paper.  Here, $\kappa=g(v)$ is the one-form dual to the Killing vector $v$. The objects $P_\pm$ are well defined projectors as they satisfy $(P_\pm )^2=P_\pm$, $P_+ P_-=0$ and $P_+ + P_- = \id$. Especially, $P_+$ induces a split of the bundle of two-forms into two orthogonal subbundles. The image of $P_+$ consists of two-forms that are self dual at the fixed points of $v$ where $\tilde s =0$, and anti-self dual at those fixed points where $s=0$. We refer to $P_+$ as a {\it flipping} projector.

Projectors $Z_\pm , \tilde Z_\pm$ analogous to $P_\pm$ can be defined for spinors on ${\mathcal M}$. Consider a Dirac spinor $\Psi$ on ${\mathcal M}$ (see appendix~\ref{app:conv} for a summary of our conventions).  Its left-handed and right-handed components are $\psi_\alpha$ and $\bar \psi^{\dot \alpha}$. Given two such spinors $\Psi_{1,2}$ we can define the $SO(4)$ invariant product $\bar \Psi_2 \Psi_1=\psi_2\psi_1+\bar \psi_2 \bar \psi_1$. The projectors $L={1\over 2}( \id + \gamma_5)$ and $R={1\over 2}( \id - \gamma_5)$ on the left- and right-handed components of $\Psi$ are compatible with the product of two spinors, that is:
\begin{equation}
  \bar \Psi_2 L \Psi_1={\overline{L \Psi_2}}\,\Psi_1~,\qquad
  \bar \Psi_2 R \Psi_1={\overline{R \Psi_2}}\,\Psi_1~.
\end{equation}
If a vector field $v$ with isolated fixed points related to $s$ and $\tilde{s}$ exists on ${\mathcal M}$, we can define a new projector acting on Dirac spinors
\begin{equation}
  \label{defZpl}
  Z_+={1\over 2}\left( \id +  {s-\tilde s \over s+\tilde s} \gamma_5
    -{2\over s+\tilde s} v^\mu \gamma_5 \gamma_\mu \right)~.
\end{equation} 
We have   $Z_+^2=Z_+$ and 
$\bar \Psi_2 Z_+ \Psi_1={\overline{Z_+ \Psi_2}}\,\Psi_1$,  the projector $Z_+$  is then compatible with the inner product. Starting from $Z_+$ we can find more projectors
\begin{equation}
  \label{defotz}
  Z_-= \id -Z_+~,\qquad
  \tilde Z_+= \gamma_5 Z_+ \gamma_5~,\qquad
  \tilde Z_-=  \id - \tilde Z_+~.
\end{equation}
The image of $Z_+$ (or $\tilde Z_+$) comprises spinors that are left-handed at the plus fixed points of $v$ and right-handed at the minus fixed points. There is a direct relation between the Killing spinors $\zeta_i$ and $\bar\chi_i$ and the projectors we introduced. Indeed we can construct a Dirac spinor ${\mathfrak z}_i$
\begin{equation}
  \label{defkildir}
  {\mathfrak z}_i=  \begin{pmatrix}
    \zeta_{i} \\
    \bar{\chi}_{i}
  \end{pmatrix}~,
\end{equation}
which satisfies $Z_+ {\mathfrak z}_i={\mathfrak z}_i$ as a consequence of~\eqref{defchi} or~\eqref{newspin}. This is a first indication of the strict relation between supersymmetry and the flipping bundles constructed here. Similarly we construct a Dirac spinor $\check {\mathfrak z}_{\check \imath}$ out of the auxiliary spinors $\check\zeta_{\check \imath},~{\check{\bar\chi}}_{\check \imath}$:
\begin{equation}
  \check {\mathfrak z}_{\check \imath}=  \begin{pmatrix}
    \check \zeta_{\check \imath} \\
    {\check{\bar \chi}}_{\check \imath}
  \end{pmatrix}~,
\end{equation}
which, using~\eqref{conchecks}, is seen to satisfy $\tilde Z_{-}\, \check {\mathfrak z}_{\check \imath}=\check {\mathfrak z}_{\check \imath}$. Let $\Psi_{1,2}$ be spinors such that $Z_+ \Psi_{1,2}=\Psi_{1,2}$. We can construct the two-form
\begin{equation}
  \omega_{\mu\nu}=\bar \Psi_2 \gamma_{\mu\nu} \Psi_1=
  \psi_2 \sigma_{\mu\nu} \psi_1+\bar \psi_2 \bar \sigma_{\mu\nu} \bar \psi_1~,
\end{equation}
which satisfies $P_+\omega=\omega$ with $P_+$ defined above in~\eqref{defpp}. This establishes a relation between the bundle of two-forms in the image of $P_+$ and the spinor bundle in the image of $Z_+$~.  In the same way, starting from two spinors such that $\tilde Z_- \Psi_{1,2}=\Psi_{1,2}$ the two-form $\bar \Psi_2 \gamma_{\mu\nu} \Psi_1$ is in the image of $P_-= \id -P_+$ (similar relations can be found using $Z_-$ or $\tilde Z_+$).

\subsection{Cohomological fields}
Here we briefly review the rewriting of the vector multiplet in terms of cohomological (or twisted) fields introduced in~\cite{Festuccia:2018rew}. We then proceed to constructing the appropriate twisted fields for the case of the hypermultiplet.

\subsubsection{Vector multiplet}
The vector multiplet comprises a complex scalar $X$, the gauge field $A$, an auxiliary scalar $D_{i j}$ and gauginos $\lambda_i,~\bar \lambda_i$.  There is an invertible map between these component fields and cohomological (twisted) fields. We present the details of the map in Appendix~\ref{app:twisvec}.  The twisted fields arrange themselves into various multiplets. There is one long multiplet made of the gauge field $A$, a scalar $\phi$, and a Grassmann one-form $\Psi$ both in the adjoint of the gauge group. Supersymmetry acts on these fields as follows:
\begin{equation}
  \label{deltalong}
  \delta \phi= \iota_v \Psi~,\quad
  \delta \Psi= \iota_v F+i d_A \phi~,\quad
  \delta A= i \Psi~.
\end{equation}
Here $F$ is the field strength of $A$ and $\iota_v$ denotes contraction with the vector field $v$.

The rest of the twisted fields arrange in two short multiplets. One is formed by a scalar $\varphi$ and a Grassmann scalar $\eta$ and the second is formed by a Grassmann two-form $\chi$ satisfying $P_+ \chi=\chi$ and a second two-form $H$ also satisfying $P_+ H=H$. All these fields are in the adjoint of the gauge group. Supersymmetry acts as follows:
\begin{equation}
  \begin{split}
    &\delta \varphi= i \eta~,\qquad\qquad
    \delta \eta= \iota_v d_A \varphi- [\phi,\varphi]~, \\
    &\delta \chi=H~,\qquad\qquad
    \delta H=i {\cal L}_v^A \chi-i [\phi,\chi]~.
  \end{split}
\end{equation}

By construction, the twisted fields above do not transform under $SU(2)_{R}$. The forms $\chi$ and $H$ are well defined only on orientable manifolds, while the latter does not need to be spin as the vector multiplet cohomological complex does not contain spinors.

With canonical reality conditions on the scalars $X^*=\bar X$, the field $\varphi$ is real. On the other hand, the reality properties of $\phi$ involve a non-trivial shift depending on $(s, \tilde s, \varphi)$:
\begin{equation}
  \phi^*= \phi+ i (s-\tilde s) \varphi~.
\end{equation}

\subsubsection{Hypermultiplet}
\label{sec:hypermultiplet}
The hypermultiplet cohomological fields are fermions $\qs, \bs, \cs, \hs$ transforming in the fundamental of $Sp(k)$. The fields $\qs$ and $\hs$ are Grassmann-even, while $\bs$ and $\cs$ are Grassmann-odd. Moreover $\qs$ and $\cs$ are in the image of the projector $Z_+$ defined in~\eqref{defZpl}, while $\bs$ and $\hs$ are in the image of $\tilde Z_-$ defined in~\eqref{defotz}:
\begin{equation}
  Z_+ \qs=\qs~,\qquad
  Z_+ \cs=\cs~,\qquad
  \tilde Z_- \bs=\bs~,\qquad
  \tilde Z_- \hs=\hs~.
\end{equation}

The fields $\qs$ and $\cs$ are related to the component fields of Section \ref{sec:gaug-hyperm} by the following map:
\begin{align}
  \qs_{n}  = {\mathfrak z}^i q_{n i}=
  \begin{pmatrix}
    \zeta^{i}q_{ni} \\
    \bar{\chi}^{i}q_{ni}
  \end{pmatrix}\quad \text{and} \quad
  \cs_n =-{s+\tilde s\over 4} Z_+ \begin{pmatrix}\psi_n \\ \bar \psi_n \end{pmatrix}
  = -\frac{1}{4}
  \begin{pmatrix}
    s\psi_{n} - v^{\mu}\sigma_{\mu}\bar{\psi}_{n} \\
    \tilde{s}\bar{\psi}_{n} + v^{\mu}\bar{\sigma}_{\mu}\psi_{n}
  \end{pmatrix}~,
\end{align} 
where we show explicitly their left- and right-handed components. The index $n$ transforms in the fundamental of $Sp(k)$.

There are two remaining cohomological fields. The first $\bs_n$ is given by:
\begin{align}
  \bs_{n}  = {s+\tilde s\over 4}\tilde Z_- \gamma_5 \begin{pmatrix} \psi_n \\ \bar \psi_n\end{pmatrix}=\frac{1}{4}
  \begin{pmatrix}
    \tilde{s}\psi_{n} + v^{\mu}\sigma_{\mu}\bar{\psi}_{n} \\
    -s\bar{\psi}_{n} + v^{\mu}\bar{\sigma}_{\mu}\psi_{n}
  \end{pmatrix}~.
\end{align}
The last field $\hs$ is related to the variation of $\bs$ under supersymmetry
$\hs=-i \,\delta \bs$:
\begin{align}
  \hs_{n} =
  {s+\tilde{s}\over 2} \check{\mathfrak z}^{\check{\imath}}F_{n \check \imath}+ \tilde Z_-\left({s+\tilde s\over 2}\gamma^\mu (D_\mu+i T_\mu) \qs_n  +{i} v^\mu G_\mu \qs_n-i  {(s+\tilde s)\over 2} {\varphi_n}^m \qs_m \right)~.
\end{align}
In this formula $T$ is a combination of supergravity background fields and derivatives of Killing spinor bilinears,
\begin{equation}
  T_\mu = {s - \tilde  s \over (s+\tilde s)} G_\mu + \frac{s  \tilde  s  }{  (s+ \tilde s )^2} b_\mu    +i  \frac{\partial_{\mu}(s^2 +  \tilde s^2)}{ 2 (s+  \tilde s  )^2} ~ ,
\end{equation}
where $b$ is a one-form satisfying $\iota_v b=0$, which parametrizes remaining freedom in choosing the supergravity background (see Appendix~\ref{sec:sugra-background-solutions}).

The map from the hypermultiplet components to the cohomological variables has a smooth inverse.  For the scalar $q_{n i}$, this is
\begin{align}
  q_{ni} =-{4\over s+\tilde s} \, \bar {\mathfrak z}_i \,\qs_n~,
\end{align}
where $\mathfrak z$ is the Dirac spinor built out of the Killing spinors defined in~\eqref{defkildir}. The map from $\bs,\cs$ to the ordinary component fields $\psi, \bar{\psi}$ is
\begin{equation}
  \begin{pmatrix}
    \psi_{n} \\
    \bar{\psi}_{n}
  \end{pmatrix}= {4\over s +\tilde s}(\gamma_5 \bs_n-\cs_n)~,
\end{equation}
and the one for the ordinary auxiliary field is
\begin{equation}
  F_{n \check{\imath}}= {8\over (s +\tilde s)^2}\left[\bar {\check{\mathfrak z}}_{\check{\imath}}\hs_{n} -{s+\tilde s\over 2}\bar {\check{\mathfrak z}}_{\check{\imath}}\gamma^\mu (D_\mu+i T_\mu) \qs_n  -{i} v^\mu G_\mu (\bar{\check{\mathfrak z}}_{\check{\imath}}\qs_n)+i  {(s+\tilde s)\over 2} {\varphi_n}^m (\bar {\check{\mathfrak z}}_{\check{\imath}}\qs_m)\right]~.
\end{equation}
Equations~\eqref{realhyp} and~\eqref{eq:sugra-reality} imply the following reality conditions on Grassmann-even spinors:
\begin{equation}
  \begin{split}
    &  \left(\hs_{n}\right)^*   = - {\bar \hs}^{n} -\left[(s+\tilde s) (D_\mu+i T_\mu) \bar \qs^n \gamma^\mu -2 i v^\mu G_\mu\bar \qs^n-i {(s+\tilde s)}  \bar\qs^m {\varphi_m}^n\right] \tilde Z_-  ~ , \\
    & \left(\qs_{n}\right)^*   = \bar \qs^{n}.
  \end{split}
\end{equation}
The twisted fields for the hypermultiplet we defined above are spinors; consequently, a spin structure is necessary in order to define them. This is different from the case of the vector multiplet whose twisted fields can be defined provided the manifold is orientable and admits a $U(1)$ action.  However, there are cases where the requirement of the manifold to be spin may be relaxed. For instance, if the theory under consideration has a $U(1)$ flavor symmetry, the twisted fields are sections of the product of the spin bundle and powers of a unitary line bundle $L$. If the manifold admits a spin$^c$ structure, such products can be well defined even if $L$ and the spin bundle do not exist. Depending on the charges under the flavor symmetry, it may be then possible to define the twisted theory on a spin$^c$ manifold (in four dimensions any closed orientable four manifold is spin$^c$). The requirement that the manifold be spin may be relaxed in other circumstances and is therefore dependent on the specific theory under consideration.

The cohomological fields we introduced split into two separate multiplets under supersymmetry:
\begin{equation}
  \begin{aligned}
    \label{eq: hccnoind}
    & \delta \qs   = \cs  ~ ,  &\qquad\qquad
    \delta \cs   =  ( i   \mathcal L_v   - \mathcal G_\Phi ) \qs  ~ , \\
    & \delta \bs  = i \hs ~ ,  &
    \delta \hs   =    (  \mathcal L_v   + i   \mathcal G_\Phi ) \bs ~ ,
  \end{aligned}
\end{equation}
where, just like in (28) of~\cite{Festuccia:2018rew}, the supersymmetry algebra encodes gauge transformations $\mathcal G_\Phi$ with respect to the field $ \Phi = ( i \iota_v A + \phi )$.  Here $A$ and $\phi$ are fields in the twisted vector multiplet (see Appendix~\ref{app:twisvec}). As in section~\ref{sec:gaug-hyperm}, the object $\mathcal G_\Phi$ acts on fields according to their representation, e.g.
\begin{equation}
  \left(\mathcal G_\Phi \, \qs \right)_n = i { \Phi_n}^m \qs_m = i \left[  i {\left(\iota_v A \right)_n}^m +   { \phi_n}^m \right] \qs_m ~ .
\end{equation}
Few comments are in order:
\begin{itemize}
\item It follows from~\eqref{deltalong} that $\mathcal G_\Phi$ is $\delta$-closed, which ensures the closure of the algebra~\eqref{eq: hccnoind}. Indeed, $\delta$ acts on the cohomological fields as an equivariant differential, namely:
  \begin{equation}
    \delta^2 = i \mathcal L_v - \mathcal G_\Phi ~ . 
  \end{equation}

\item The action of supersymmetry commutes with the projectors $Z_{\pm},\tilde Z_\pm$. This is because the vector field that enters in their definition is the same as the Killing vector that appears in the supersymmetry variations (in general it is sufficient for the two vector fields to commute).

\item The gauge field $A$ and the scalar field $\phi$ always appear in the combination $\Phi $. When discussing localization we will see that, after gauge fixing, $\Phi$ will be a Coulomb branch modulus appearing in one-loop determinants. If $\Phi$ is part of a dynamical vector multiplet, it will be integrated over. $\Phi$ could also arise from a background vector multiplet, in which case it will remain as a free parameter in supersymmetric observables.  For instance it can be identified with a flavour fugacity.
\end{itemize}

\subsection{Hypermultiplet Lagrangian in cohomological fields}
We have shown in section~\ref{sec:gaug-hyperm} that the usual quadratic Lagrangian for a hypermultiplet coupled to a rigid ${\cal N}=2$ supergravity background is the $\delta$ variation of $V_{G}$ in~\eqref{eq:qexact_lagr}. In terms of the cohomological fields $( \mathfrak q , \mathfrak b , \mathfrak c , \mathfrak h )$, the deformation term $V_{G}$ reads as follows:
\begin{align}
  V_{G} =
  \frac{8}{(s+\tilde{s})^{3}}\Bigg{\{}
  & i \bar \cs \mathcal{L}_v\qs
    +i \bar \qs\left[\phi +i (s-\tilde{s})\varphi \right]\cs
    -{i} (\partial_{\mu}v_{\nu}){\bar \cs}\,\gamma^{\mu\nu}\qs  \nonumber \\
  &-(s+\tilde{s})\left(G_{\mu} - \frac{s^{2}-\tilde{s}^{2}}{64}b_{\mu}\right){\bar \cs} \gamma^{\mu} \qs
    +i(s+\tilde{s})\bar \bs \gamma^{\mu}(D_{\mu}+i T_\mu)\qs \nonumber \\
  & -(s+\tilde{s}) {\bar \bs}\, \varphi\, \qs
    - 2 \iota_v G {\bar \bs} \qs
    - \frac{i}{4}(s+\tilde{s})^{2}{\bar \qs} \chi \qs
    -i  {\bar\hs} \bs
    \Bigg{\}}~ , 
\end{align}
where, in the expression above, $Sp(k)$ gauge/flavor indices are contracted  as $\overline \Psi_1 \Psi_2 = (\overline \Psi_1)_m  \Psi_2^m$. Here we used the  short-hand notation:
\begin{equation}
  \chi = \frac{1}{2} \chi_{\mu\nu} \gamma^{\mu\nu}~.
\end{equation}
By taking a $\delta$ variation of $V_G$, we obtain the Lagrangian for the hypermultiplet ${\cal L}={\cal L}_B+{\cal L}_F$:
\begin{align}
  \mathcal{L}_{B} =
  \frac{8}{(s+\tilde{s})^{3}}\Bigg{\{}
  &-\mathcal{L}_v \bar \qs \mathcal{L}_v\qs
    - D_{\mu}v_{\nu}\bar \qs \gamma^{\mu\nu}\mathcal{L}_v\qs
    -i(s-\tilde{s})\bar \qs\, \varphi\, \mathcal{L}_v\qs \nonumber\\
  &+i(s+\tilde{s})\left(G_{\mu} - \frac{s^{2}-\tilde{s}^{2}}{64}b_{\mu}\right)( \bar \qs \gamma^{\mu}\mathcal{L}_v\qs +  \bar \qs \gamma^{\mu} \phi\, \qs)\nonumber\\
  &- \partial_{\mu}v_{\nu}\bar \qs \gamma^{\mu\nu}\phi\, \qs
    - \frac{1}{2}\bar \qs \left[ \frac{i}{2}(s+\tilde{s})^{2} H^{+}
    + \{\phi,\phi\}
    + i(s-\tilde{s})\{\varphi,\phi\}  \right]\qs \nonumber\\
  & -(s+\tilde{s})\bar \hs \gamma^{\mu}(D_{\mu}+i T_\mu)\qs
    - 2i  \iota_v G \bar \hs \qs
    + i (s+\tilde{s}) \bar \qs \,\varphi\, \hs 
    +\bar \hs \hs
    \Bigg{\}}~,
\end{align}
and
\begin{align}
  \mathcal{L}_F =
   \frac{8}{(s+\tilde{s})^{3}}\Bigg{\{}
  &-i  \bar \bs \mathcal{L}_v\bs +i  \bar \bs \phi \bs
    -i \bar \cs \mathcal{L}_v\cs
    -i(s+\tilde{s})\bar \bs \gamma^{\mu}(D_{\mu}+iT_\mu)\cs
        + 2 \iota_v G \bar \bs \cs \nonumber\\
  &+(s+\tilde{s})\left(G_{\mu} - \frac{s^{2}-\tilde{s}^{2}}{64}b_{\mu}\right) \bar \cs\, \sigma^{\mu}\cs
    +i \bar \cs \left[ \phi +i (s-\tilde{s})\varphi \right]\cs
    +i \partial_{\mu}v_{\nu}\bar \cs\,\sigma^{\mu\nu}\cs  \nonumber\\
  &+  (s+\tilde{s})\bar \cs\,\varphi\, \bs
    + i(s+\tilde{s})\bar \qs \gamma^\mu \Psi_\mu \bs
    +i (s+\tilde{s})\bar \qs\, \eta\, \bs  \nonumber\\
  &+ i\bar\qs \left[  2\iota_v \Psi
    -(s-\tilde{s})\eta
    +\frac{1}{2}(s+\tilde{s})^{2}\chi \right]\cs
    \Bigg{\}}~.
\end{align}

\subsection{Non-equivariant topological twist}
In our setup, pure topological twisting is recovered by turning off either $(\bar \chi_{i \dot\alpha} , \check \zeta_{i \alpha} )$ or $(\zeta_{i \alpha} , \check {\bar \chi}_{i \dot\alpha} )$. In terms of Killing spinor bilinears, this corresponds to respectively setting either $\tilde s = 0 $ and $ s = {\rm constant} $ or $ \tilde s = {\rm constant} $ and $ s = 0 $. The Killing vector $v$ vanishes in both cases, yielding a non-equivariant differential $\delta$ satisfying $\delta^2=0$. In turn, the hypermultiplet cohomological complex fulfills $\delta \mathfrak h = \delta \mathfrak c =0$, while Dirac spinors become pure Weyl spinors:
\begin{align}
  & ( \mathfrak q , \mathfrak b , \mathfrak c , \mathfrak h  )_{\tilde s = 0} = ( L \, \mathfrak q , R \, \mathfrak b , L \, \mathfrak c , R \,  \mathfrak h  ) ~ ,    & ( \mathfrak q , \mathfrak b , \mathfrak c , \mathfrak h  )_{  s = 0} = ( R \, \mathfrak q , L \, \mathfrak b , R \, \mathfrak c , L \,  \mathfrak h  ) ~ .  
\end{align}
Consequently, the deformation term $V_G$ becomes 
\begin{equation}
  \begin{split}
    & \mathcal V_G =  V_G |_{ \widetilde s = 0}  =  \frac{8}{ s^{2}} [  \bar \bs ( i \gamma^{\mu}  D_{\mu} - \varphi  )\qs   - \frac{i \, s}{4}  {\bar \qs} \chi \qs  -i  {\bar\hs} \bs  ]  ~ , \\
    & \tilde{\mathcal V}_G =  V_G |_{ s = 0}  =  \frac{8}{ \tilde{s}^{2}} [  \bar \bs ( i \gamma^{\mu}  D_{\mu} - \varphi  )\qs   - \frac{i \, \tilde{s} }{4} {\bar \qs} \chi \qs  -i  {\bar\hs} \bs  ] ~ .
  \end{split}
\end{equation}
The objects ${\mathcal V}_G$ and $\tilde{\mathcal V}_G$ are  deformation terms reproducing the features of topologically twisted hypermultiplets, see for instance~\cite{Labastida:2005zz} and references therein.

\section{Cohomological localization}\label{sec:5}
In this section we perform supersymmetric localization in the cohomological language described previously. The localization argument is the standard one: let $Z_{\rm hyper}$ be the hypermultiplet partition function, which is a path integral with functional weights $e^{ - S_{\rm hyper}}$.  We then make the replacement $S_{\rm hyper} \to S_{\rm hyper} + t \, S_{\rm loc} $, with $t$ being a real parameter and $S_{\rm loc} $ a $\delta$-exact positive definite deformation action.  The resulting partition function $Z(t)$ naively depends on $t$ and is such that $Z(0)=Z_{\rm hyper}$. In fact, $Z(t)$ is independent of $t$ because both the path integral measure and the integrand are invariant under $\delta$. This implies that $ Z(\infty) = Z_{\rm hyper}$, meaning that the latter can be computed exactly via saddle point method by using the integrand $e^{ - S_{\rm loc}}$. Consequently, $ Z_{\rm hyper}$ is given by $e^{ - S_{\rm loc}}$ evaluated on the locus of BPS field configurations, times the corresponding 1-loop determinant. Actually, $ Z_{\rm hyper}$ can elegantly be obtained from the index of a transversally elliptic operator, as we shall see. After gauge fixing, the supersymmetry transformation $\delta$ becomes an equivariant differential squaring to $\delta^2 = i \mathcal L_v - \mathcal G_{a_0}$, where $a_0$ is a constant valued in the Cartan subalgebra of the gauge group $G$. A priori, then, $ Z_{\rm hyper}$ will be a function of $a_0$. If the manifold admits non-trivial fluxes with magnetic charges $\mathfrak m_i$, the value of the Coulomb branch parameter $a_0$ is suitably shifted by $\mathfrak m_i$, as in the case of $ S^2 \times S^2$~\cite{Bawane:2014uka}.

\subsection{BPS Locus}     
On the BPS locus, the fermionic fields $\mathfrak b , \mathfrak c $ and their supersymmetric variations are vanishing. In particular, this yields $ \hs = 0$ and
\begin{align}
  & \left(i \mathcal L_v - \mathcal G_\Phi \right) \qs=0 ~ .
\end{align}
Since the field $\Phi$ is complex and the Killing vector $v$ is real, by imposing reality conditions on the above equations we end up with a trivial BPS locus: $\mathfrak q = 0$.

\subsection{One-loop determinant from index theorem}
The deformation term we employ is
\begin{align}
  & V_{\rm loc} =  \frac1{4 \, \overline{  \mathfrak z}^i \mathfrak z_i  }  { (\delta \Psi_n) }^* \Psi_n =  \frac8{ (s+\tilde s)^3 }  [ (\delta \mathfrak b_n) ^* \mathfrak b_n +  (\delta \mathfrak c_n) ^* \mathfrak c_n ] ~ .
\end{align}
By construction, $V_{\rm loc} $ leads to a positive definite deformation
Lagrangian $\mathcal L_{\rm loc} = \delta V_{\rm loc} $. Explicitly,
\begin{align}
  V_{\rm loc} 
  & =  \frac8{(s+\tilde s)^3} \Big\{   -  \overline{  \mathfrak b  } \delta \mathfrak b + \overline{\mathfrak b} \left[ i ( s + \tilde s)  \gamma^\mu (D_\mu + i T_\mu)  - 2  \iota_v G    - g (s + \tilde s)  \varphi      \right] \mathfrak q    \nonumber \\ 
  &\hspace{140pt}  +  (\overline{\delta  \mathfrak q})  \left[   i \mathcal L_v  - i \Phi  -  i  ( 2\phi + i(s-\tilde s)\varphi ) \right] \mathfrak q   \Big\} ~ .  
\end{align}
The deformation term $V_{\rm loc}$ can be recast in the quadratic form
\begin{align}
  V_{\rm loc} = \frac{ 8  } { (s + \tilde s)^3 }  \begin{pmatrix}  \overline{ \delta\qs }   , \,\,  \overline{ \bs}   \end{pmatrix} \begin{pmatrix}  D_{00} &&  D_{01}  \\  D_{10}  &&  D_{11} \end{pmatrix}      \begin{pmatrix}  \qs    \\  \delta\bs  \end{pmatrix}  ~ , 
\end{align}
with entries 
\begin{align}
  \begin{aligned}
    D_{00} &=   i \mathcal L_v  - i \Phi  -  i  [  2\phi + i(s-\tilde s)\varphi ] ~ , \\
    D_{10} &= i ( s + \tilde s)  \gamma^\mu (D_\mu + i T_\mu)  - 2  \iota_v G    - g (s + \tilde s)  \varphi    ~ ,
  \end{aligned}
           &&
  \begin{aligned}
    D_{01} &=   0   ~ , \\
    D_{11} &= - \id    ~ .
  \end{aligned}
\end{align}  
The operator $D_{10}$ implicitly fulfills $D_{10} = \tilde Z_- D_{10} Z_+ $ because it acts on $\mathfrak q$ and $\mathfrak b$, which are Dirac spinors satisfying the projection conditions $ Z_+ \mathfrak q = \mathfrak q$ and $ \widetilde Z_- \mathfrak b = \mathfrak b$. Furthermore, $ D_{10} $ is a transversally elliptic operator. Ellipticity of a differential operator $\mathscr D$ amounts to the invertibility of the corresponding symbol $\sigma[\mathscr D]$, where the latter is obtained by keeping only the highest derivative terms in $ \mathscr D $ and making the substitution $\partial_\mu \to i p_\mu$. If $\sigma[\mathscr D]$ is invertible for any $p^\mu\neq0$, then $\mathscr D$ is elliptic. For instance, the Dirac operator $\mathscr D = - i \gamma^\mu \partial_\mu$ has symbol $\sigma[\mathscr D] = \gamma^\mu p_\mu $, which is everywhere invertible in flat Euclidean space.  As for $ D_{10}$, its symbol is
\begin{align}
  \sigma\left[   D_{10} \right]  =   \frac{ 8 \,  p_\mu }{ ( s + \tilde s )^2}  \, \tilde{  Z}_- \gamma^\mu  Z_+  ~  .
\end{align}
At plus fixed points we have 
\begin{align}
  &  \left. \sigma\left[   D_{10} \right]  \right|_{\tilde s=0, v=0}  =   \frac{ 8 \,  p_\mu }{  s^2}  \, R \gamma^\mu  L  \to \frac{ 8 \,  p_\mu }{  s^2}  \,  \overline \sigma^\mu =  \sigma\left[ - \frac{8 i}{ s^2} \overline \sigma^\mu \partial_\mu \right]     ~  ,  
\end{align}
while at minus fixed points we find
\begin{align}
  & \left. \sigma\left[   D_{10} \right]  \right|_{  s=0, v=0}  =   \frac{ 8 \,  p_\mu }{  \tilde s^2}  \, L \gamma^\mu  R \to   \frac{ 8 \,  p_\mu }{  \tilde s^2}  \,   \sigma^\mu  =   \sigma\left[  - \frac{8 i}{ \tilde s^2}    \sigma^\mu \partial_\mu \right]    ~  .
\end{align}
Consequently, the symbol $ \sigma\left[ D_{10} \right] $ coincides with that of a chiral Dirac operator whenever $ \tilde s = 0 $ or $ s = 0$, ensuring ellipticity of $D_{10}$. On the other hand,
\begin{align}
  & \left. \sigma\left[   D_{10} \right]  \right|_{  s = \tilde s }  =    \frac{    p_\mu }{ 2 s^2}  \, \left( 1 - \frac{v_\nu}s \gamma_5 \gamma^\nu  \right)  \gamma^\mu  \left( 1 - \frac{v_\rho}s \gamma_5 \gamma^\rho  \right)  = \frac{   4  }{ s^2}  \, \widetilde{ Z}_-\gamma_5 \gamma^{\mu\nu} p_\mu v_\nu  ~  , 
\end{align}
implying that $ D_{10} $ is not elliptic in the patch where $s = \tilde s$, because $\left. \sigma\left[ D_{10} \right] \right|_{ s = \tilde s } = 0 $ for $p^\mu = v^\mu \neq 0$. The operator $ D_{10}$ is then transversally elliptic with respect to the Killing vector $v$, as the symbol $\sigma\left[ D_{10} \right] $ is everywhere invertible for any $p^\mu\neq0$ such that $p\cdot v = 0$.
 
The hypermultiplet contribution to the partition function is encoded into the index of the transversally elliptic operator $ D_{10}$~\cite{Pestun:2007rz,Hama:2012bg}:
\begin{align}\label{eq: indexformula}
  & {\rm ind}(   D_{10})(t) = \sum_{ x \, : \,  \widetilde x = x} \frac{ {\rm Tr}_\qs e^{ - i \, t \, \mathcal H } - {\rm Tr}_\bs e^{ - i \, t \, \mathcal H }  }{\det(1-\partial \widetilde x / \partial x)} ~ , & \mathcal H = \delta^2 = i \mathcal L_v - \mathcal G_\Phi  ~ ,
\end{align}
where $t \in \mathbb R$, while $\widetilde x$ is the image of the coordinates $x$ under the torus action induced by $\mathcal H$ and the sum is over the fixed points $\widetilde x = x$. In a neighborhood of a plus fixed point, the metric becomes flat and the manifold can be parametrized by a pair of complex coordinates $ ( z_1 , z_2 ) $. In (\ref{eq: indexformula}), bosons $\mathfrak q$ contributes to ${\rm ind}(   D_{10})(t)$ with a plus sign, whereas fermions $\mathfrak b$ contributes with a minus sign.

The Killing vector $v$ in a neighborhood of $\tilde s = 0$ reads
\begin{equation}\label{eq: locppkv}
  v = i \epsilon_1^{(+)}  (z_1 \partial_{z_1} - \bar z_1 \partial_{\bar z_1} ) + i \epsilon_2^{(+)}  (z_2 \partial_{z_2} - \bar z_2 \partial_{\bar z_2} ) ~ ,
\end{equation}
with $  \epsilon_1^{(+)} $ and $  \epsilon_2^{(+)} $ being real parameters. Hence,   $ e^{ - i \, t \, \mathcal H } $ is a $U(1)\times U(1)$ action attaching to $(z_1 , z_2)$   phases $q_i$ with $i=1,2$:
\begin{align}
  z_i \to \widetilde z_i =  q_i z_i ~ ,  \qquad  q_i = \exp( i \, \epsilon_i^{(+)} \, t ) ~ , \qquad  i = 1,2 ~ .
\end{align}
In a neighborhood of a plus point, the denominator entering the index formula (\ref{eq: indexformula}) is
\begin{equation}
  \det\left(1 - \frac{\partial \widetilde z_i}{\partial  z_j } \right) = (1-q_1)(1-\overline q_1 ) (1-q_2)(1- \overline q_2 ) ~ ,
\end{equation}
where $\overline q_i = q_i^{-1}$ is the complex conjugate/inverse of the phase $q_i$. We now need the action of $ \mathcal H $ upon the spinors $\mathfrak q$ and $\mathfrak b$.  We then embed $U(1)\times U(1)$ into $SU(2)_+ \times SU(2)_- \sim \text{Spin}(4)$ to see how spinors transform under $ \mathcal H $. We define
\begin{align}
  &{ \bf z}   = x_\mu \gamma^\mu  =    \begin{pmatrix}  
    0 && 0 && \overline z_2 &&   \overline z_1   \\  0  && 0 &&      z_1   &&   - z_2  \\
    - z_2 && - \overline z_1 && 0 && 0 \\
    - z_1 &&  \overline z_2 && 0 && 0 \\
  \end{pmatrix}   ~ .
\end{align}
Then, coordinates ${ \bf z} $ and spinors $\Psi = \{ \mathfrak q , \mathfrak b \} $ transform as
\begin{align}
  &     { \bf z}  \to  g \,  { \bf z}  \,   g^{-1} ~ , \qquad     { \Psi }  \to     g^{-1} \Psi ~ ,   & g = \text{diag}\left( \sqrt{\overline q_1 \overline q_2}  ,  \sqrt{ q_1 q_2} ,  \sqrt{ \overline  q_1 q_2  } , \sqrt{ q_1 \overline q_2 }  \right) ~ ,   
\end{align}
with $g \in SU(2)_+ \times SU(2)_-$ being the spinor representation of the torus action $\mathcal L_v$. At a plus point, $\mathfrak q$ is left-handed and $\mathfrak b$ is right-handed: $\mathfrak q = L \, \mathfrak q$ and $\mathfrak b = R \, \mathfrak b$. Then, the action of $ \mathcal L_v $ upon $\mathfrak q , \mathfrak b $ is
\begin{align}
  &  \mathfrak q_+ \to  \sqrt{  q_1   q_2} \,  \mathfrak q_+ ~ , \qquad   \mathfrak q_- \to  \sqrt{ \overline  q_1 \overline  q_2} \,  \mathfrak q_- ~ , \qquad   \widetilde{\mathfrak b }^{ \dot +} \to  \sqrt{ q_1 \overline q_2 } \,  \widetilde{\mathfrak b }^{ \dot +} ~ , \qquad   \widetilde{ \mathfrak b}^{\dot -} \to  \sqrt{  \overline q_1  q_2} \,   \widetilde{ \mathfrak b}^{\dot -}  ~ .  
\end{align}
The operator $\mathcal H$ also contains $\mathcal G_\Phi$, which acts non-trivially on $\mathfrak q , \mathfrak b$. Thus, for a hypermultiplet in the representation $\mathcal R $ of the gauge group, the index formula (\ref{eq: indexformula}) yields
\begin{equation}
  \left.{\rm ind}(  D_{10})\right|_{\text{plus point}} = \frac{\sqrt{q_1 q_2}}{(1-q_1)(1-q_2)} \sum_{\rho \in \mathcal  R} e^{ -  t \, \rho(\Phi_0)   }   ~ .
\end{equation}
This index translates into a functional determinant by   expanding ${\rm ind}(  D_{10})|_{\text{plus point}} $ in powers of $q_1, q_2$ and converting the corresponding series into an infinite product. The latter requires regularization, which is  a delicate matter. For instance, in cases linked to   five-dimensional manifolds, the regularization was established in~\cite{Qiu:2013pta,Qiu:2013aga,Qiu:2014cha,Qiu:2014oqa,Festuccia:2016gul}.    Here, we will examine   diverse regularizations. 

We define:
\begin{align}
& \left[ \frac1{1-q_i} \right]_+ = \sum_{n\geq0} q_i^n ~ , \qquad   \left[ \frac1{1-q_i} \right]_-   = - \sum_{n\leq-1}   q_i^{n} = - \sum_{n \geq 0}   q_i^{ - n - 1} ~ .
\end{align}
The difference between the two regularizations is 
\begin{equation}
  \left[ \frac1{1-q_i} \right]_+ -   \left[ \frac1{1-q_i} \right]_- = \sum_{n\in \mathbb Z} q_i^n  ~ ,
\end{equation}
which  becomes a periodic Dirac delta supported at $t=0$ if $q_i = e^{i t}$.   Whenever both $q_1$ and $q_2$ appear, we write $[ \dots ]_{\pm \pm}$, where the first (respectively, the second) subscript refers to the regularization of $q_1$ ($q_2$). Indeed, the index ${\rm ind}(  D_{10})|_{\text{plus point}} $ depends on both $q_1$ and $q_2$, and we have four possible series expansions: 
\begin{equation}
  \begin{split}
    &  \left[ \left.{\rm ind}(  D_{10})\right|_{\text{plus point}} \right]_{++}   = +  \sum_{\rho \in \mathcal  R} \sum_{ n_1 , n_2 \in \mathbb N } q_1^{n_1 + \frac12 } q_2^{n_2 + \frac12 }  e^{ -  t \, \rho(\Phi_0)   }  ~ ,  \\
    &  \left[ \left.{\rm ind}(  D_{10})\right|_{\text{plus point}} \right]_{+-}   = -  \sum_{\rho \in \mathcal  R} \sum_{ n_1 , n_2 \in \mathbb N } q_1^{n_1 + \frac12 } q_2^{ - n_2 - \frac12 }  e^{ -  t \, \rho(\Phi_0)   }  ~ ,  \\
    &  \left[ \left.{\rm ind}(  D_{10})\right|_{\text{plus point}} \right]_{-+}   = -  \sum_{\rho \in \mathcal  R} \sum_{ n_1 , n_2 \in \mathbb N } q_1^{ - n_1 -  \frac12 } q_2^{n_2 + \frac12 }  e^{ -  t \, \rho(\Phi_0)   }  ~ ,  \\
    &  \left[ \left.{\rm ind}(  D_{10})\right|_{\text{plus point}} \right]_{--}   = +  \sum_{\rho \in \mathcal  R} \sum_{ n_1 , n_2 \in \mathbb N } q_1^{ - n_1 -  \frac12 } q_2^{ - n_2 - \frac12 }  e^{ -  t \, \rho(\Phi_0)   }  ~ .
  \end{split}
\end{equation}
 Here, $\rho$ is a sum over weights in the representation $\mathcal R$ and $\Phi_0$ is
\begin{equation}
  \Phi_0 =   a_0  +  k_+( \epsilon_1^{(+)}  ,  \epsilon_2^{(+)}  ) ~ ,
\end{equation}
with $a_0$ being a Coulomb branch moduli and $k_+( \epsilon_1^{(+)} , \epsilon_2^{(+)} ) $ parametrizing the flux contribution at the plus fixed point. As in~\cite{Hama:2012bg,Festuccia:2018rew}, the index $ {\rm ind}( D_{10}) |_{\text{plus point}} $ translates into a 1-loop determinant given by an infinite product. For each regularization, we have 
\begin{align}
 \left[ Z^\text{HM}_{\epsilon^{(+)}_1 , \epsilon^{(+)}_2}( a_0 , k_+ ) \right]_{++}
  &   = \prod_{\rho \in \mathcal R} \prod_{n_1 , n_2 \in \mathbb N}\left[  \epsilon^{(+)}_1 \left(n_1 + \frac12 \right) + \epsilon^{(+)}_2 \left (n_2 + \frac12 \right) + i \, \rho(\Phi_0) \right]^{-1} ~ , \nonumber \\
  & = \prod_{\rho \in \mathcal R} \Gamma_2 ( i \, \rho(\Phi_0) + ( (\epsilon^{(+)}_1 + \epsilon^{(+)}_2)/2 )  |  \epsilon^{(+)}_1 , \epsilon^{(+)}_2 )  ~ ,  
\end{align}
as well as 
\begin{equation}
  \begin{split}
    \left[ Z^\text{HM}_{\epsilon^{(+)}_1 , \epsilon^{(+)}_2}( a_0 , k_+ ) \right]_{+-}
    &   = \prod_{\rho \in \mathcal R} \prod_{n_1 , n_2 \in \mathbb N}\left[  \epsilon^{(+)}_1 \left(n_1 + \frac12 \right) - \epsilon^{(+)}_2 \left (n_2 + \frac12 \right) + i \, \rho(\Phi_0) \right] ~ ,  \\
    & =   \prod_{\rho \in \mathcal R} \left[ \Gamma_2 ( i \, \rho(\Phi_0) + ( (\epsilon^{(+)}_1 - \epsilon^{(+)}_2)/2 )  |  \epsilon^{(+)}_1 ,  - \epsilon^{(+)}_2 ) \right]^{-1}  ~ ,   \\
    \left[ Z^\text{HM}_{\epsilon^{(+)}_1 , \epsilon^{(+)}_2}( a_0 , k_+ ) \right]_{-+}
    &   = \prod_{\rho \in \mathcal R} \prod_{n_1 , n_2 \in \mathbb N}\left[  -  \epsilon^{(+)}_1 \left(n_1 + \frac12 \right) + \epsilon^{(+)}_2 \left (n_2 + \frac12 \right) + i \, \rho(\Phi_0) \right] ~ ,  \\
    & =   \prod_{\rho \in \mathcal R} \left[ \Gamma_2 ( i \, \rho(\Phi_0) + (  ( - \epsilon^{(+)}_1 + \epsilon^{(+)}_2)/2 )  | - \epsilon^{(+)}_1 ,   \epsilon^{(+)}_2 ) \right]^{-1}  ~ ,  \\
    \left[ Z^\text{HM}_{\epsilon^{(+)}_1 , \epsilon^{(+)}_2}( a_0 , k_+ ) \right]_{--}
    &   = \prod_{\rho \in \mathcal R} \prod_{n_1 , n_2 \in \mathbb N}\left[  -  \epsilon^{(+)}_1 \left(n_1 + \frac12 \right) - \epsilon^{(+)}_2 \left (n_2 + \frac12 \right) + i \, \rho(\Phi_0) \right]^{-1} ~ ,  \\
    & =   \prod_{\rho \in \mathcal R} \left[ \Gamma_2 ( i \, \rho(\Phi_0) - (   (  \epsilon^{(+)}_1 + \epsilon^{(+)}_2)/2 )  | - \epsilon^{(+)}_1 ,  - \epsilon^{(+)}_2 ) \right]  ~ .  
  \end{split}
\end{equation}
These are  1-loop contributions to the hypermultiplet partition function at a plus fixed point. We used Barnes multiple zeta and gamma functions~\cite{FRIEDMAN2004362,Felder_2000},
\begin{align}
  & \zeta_N( s , \omega | \vec a ) = \sum_{\vec n \in \mathbb N^N}(\omega + \vec a \cdot \vec m )^{-s} ~ , & \Gamma_N(   \omega | \vec a ) = \prod_{\vec n \in \mathbb N^N}(\omega + \vec a \cdot \vec m )^{-1} = e^{\partial_s \zeta_N( s , \omega | \vec a )|_{s=0} } ~ .
\end{align}
to regularize the infinite product that gives $ Z^\text{HM}_{\epsilon^{(+)}_1 , \epsilon^{(+)}_2}( a_0 , k_+ ) $. For instance, in the case of squashed $S^4$ worked out in~\cite{Hama:2012bg}, the regularization chosen for the plus point contribution is $++$: 
\begin{equation}
 Z^\text{HM}_{\epsilon^{(+)}_1 , \epsilon^{(+)}_2}( a_0 , k_+ )   \equiv  \left[ Z^\text{HM}_{\epsilon^{(+)}_1 , \epsilon^{(+)}_2}( a_0 , k_+ ) \right]_{++} ~ . 
 \end{equation}
Similarly, in a neighborhood of a minus fixed point the manifold can be parametrized by a pair of complex coordinates $ ( z_1' , z_2' ) $ and the Killing vector $v$ in a neighborhood of $ s = 0$ is
\begin{equation}\label{eq: locmpkv}
  v = i \epsilon_1^{(-)}  (z_1' \partial_{z_1'} - \bar z_1' \partial_{\bar z_1'} ) + i \epsilon_2^{(-)}  (z_2' \partial_{z_2'} - \bar z_2' \partial_{\bar z_2'} ) ~ ,
\end{equation}
with $ \epsilon_1^{(-)} $ and $ \epsilon_2^{(-)} $ being real parameters. The $U(1)\times U(1)$ action of $v$ upon $(z_1' , z_2' )$ reads
\begin{align}
  z_i' \to \widetilde z_i' =  q_i' z_i' ~ ,  \qquad  q_i' = \exp( i \, \epsilon_i^{(-)} \, t ) ~ , \qquad  i = 1,2 ~ .
\end{align}
In a neighborhood of a minus point, the denominator entering the index formula (\ref{eq: indexformula}) is formally the same as the one computed at the plus point:
\begin{equation}
  \det\left(1 - \frac{\partial \widetilde z_i'}{\partial  z_j' } \right) = (1-q_1')(1-\overline q_1' ) (1-q_2' )(1- \overline q_2' ) ~ ,
\end{equation}
At $s=0$, there is a chirality flip with respect to the plus fixed point at $\tilde s=0$. Indeed, $\mathfrak q$ is right-handed and $\mathfrak b$ is left-handed: $\mathfrak q = R \, \mathfrak q$ and $\mathfrak b = L \, \mathfrak b$. Therefore,
\begin{align}
  &  \mathfrak b_+ \to  \sqrt{  q_1'   q_2' } \,  \mathfrak b_+ ~ , \qquad   \mathfrak b_- \to  \sqrt{ \overline  q_1'  \overline  q_2' } \,  \mathfrak b_- ~ , \qquad   \tilde{\mathfrak q }^{ \dot +} \to  \sqrt{ q_1' \overline q_2' } \,  \tilde{\mathfrak q }^{ \dot +} ~ , \qquad   \widetilde{ \mathfrak q}^{\dot -} \to  \sqrt{  \overline q_1'  q_2' } \,   \widetilde{ \mathfrak q}^{\dot -}  ~ .
\end{align}
By taking into account the action of $\mathcal G_{\Phi'}$ at a minus fixed point, the index formula (\ref{eq: indexformula}) provides
\begin{align}
  \left.{\rm ind}(  D_{10})\right|_{\text{minus point}}
  & =  - \frac{\sqrt{q_1'  q_2' }}{(1-q_1')(1 -   q_2')} \sum_{\rho \in  \mathcal R}  e^{ - t \,  \rho ( \Phi_0'  ) } ~ .
\end{align}
Again, we have four possible series expansions:
\begin{equation}
  \begin{split}
    \left[ \left.{\rm ind}(  D_{10})\right|_{\text{minus point}} \right]_{++}
    &  =   - \sum_{\rho \in  \mathcal R} \sum_{n_1, n_2 \in \mathbb N}  (q_1')^{  n_1 + \frac12 } (  q_2')^{ n_2 + \frac12 } e^{ - t \,  \rho ( \Phi_0'  ) }   ~ ,  \\
    \left[ \left.{\rm ind}(  D_{10})\right|_{\text{minus point}} \right]_{+-}
    &  = +   \sum_{\rho \in  \mathcal R} \sum_{n_1, n_2 \in \mathbb N}  (q_1')^{  n_1 + \frac12 } (  q_2')^{ - n_2 - \frac12 } e^{ - t \,  \rho ( \Phi_0'  ) }   ~ ,  \\
    \left[ \left.{\rm ind}(  D_{10})\right|_{\text{minus point}} \right]_{-+}
    &  = +   \sum_{\rho \in  \mathcal R} \sum_{n_1, n_2 \in \mathbb N}  (q_1')^{ -  n_1 - \frac12 } (  q_2')^{ n_2 + \frac12 } e^{ - t \,  \rho ( \Phi_0'  ) }   ~ ,  \\
    \left[ \left.{\rm ind}(  D_{10})\right|_{\text{minus point}} \right]_{--}
    &  =   - \sum_{\rho \in  \mathcal R} \sum_{n_1, n_2 \in \mathbb N}  (q_1')^{  - n_1 - \frac12 } (  q_2')^{ - n_2 - \frac12 } e^{ - t \,  \rho ( \Phi_0'  ) }   ~ , 
  \end{split}
\end{equation}
where $\Phi_0'$ is
\begin{equation}
  \Phi_0' =  a_0'  +  k_-( \epsilon_1^{(-)}  ,  \epsilon_2^{(-)}  ) ~ ,
\end{equation}
with $k_-( \epsilon_1^{(-)} , \epsilon_2^{(-)} ) $ encoding the flux contribution at the minus fixed point.  We now translate the index $ {\rm ind}( D_{10}) |_{\text{minus point}} $ into 1-loop determinants:
\begin{equation}
  \begin{split}
 \left[ Z^\text{HM}_{\epsilon^{(-)}_1 , \epsilon^{(-)}_2}( a_0' , k_- ) \right]_{++}
  &   = \prod_{\rho \in \mathcal R} \prod_{n_1 , n_2 \in \mathbb N}\left[  \epsilon^{(-)}_1 \left(n_1 + \frac12 \right) + \epsilon^{(-)}_2 \left (n_2 + \frac12 \right) + i \, \rho(\Phi_0') \right] ~ ,  \\
  & = \prod_{\rho \in \mathcal R} \left[ \Gamma_2 ( i \, \rho(\Phi_0') + ( (\epsilon^{(-)}_1 + \epsilon^{(-)}_2)/2 )  |  \epsilon^{(-)}_1 , \epsilon^{(-)}_2 ) \right]^{-1}  ~ ,    \\
 \left[ Z^\text{HM}_{\epsilon^{(-)}_1 , \epsilon^{(-)}_2}( a_0' , k_- ) \right]_{+-}
  &   = \prod_{\rho \in \mathcal R} \prod_{n_1 , n_2 \in \mathbb N}\left[  \epsilon^{(-)}_1 \left(n_1 + \frac12 \right) - \epsilon^{(-)}_2 \left (n_2 + \frac12 \right) + i \, \rho(\Phi_0') \right]^{-1}  ~ ,  \\
  & = \prod_{\rho \in \mathcal R} \left[ \Gamma_2 ( i \, \rho(\Phi_0') + ( (\epsilon^{(-)}_1 - \epsilon^{(-)}_2)/2 )  |  \epsilon^{(-)}_1 , - \epsilon^{(-)}_2 ) \right] ~ ,    \\
 \left[ Z^\text{HM}_{\epsilon^{(-)}_1 , \epsilon^{(-)}_2}( a_0' , k_- ) \right]_{-+}
  &   = \prod_{\rho \in \mathcal R} \prod_{n_1 , n_2 \in \mathbb N}\left[  -  \epsilon^{(-)}_1 \left(n_1 + \frac12 \right) + \epsilon^{(-)}_2 \left (n_2 + \frac12 \right) + i \, \rho(\Phi_0') \right]^{-1} ~ ,  \\
  & = \prod_{\rho \in \mathcal R} \left[ \Gamma_2 ( i \, \rho(\Phi_0') + ( ( - \epsilon^{(-)}_1 + \epsilon^{(-)}_2)/2 )  | -  \epsilon^{(-)}_1 , \epsilon^{(-)}_2 ) \right]  ~ ,    \\
   \left[ Z^\text{HM}_{\epsilon^{(-)}_1 , \epsilon^{(-)}_2}( a_0' , k_- ) \right]_{--}
  &   = \prod_{\rho \in \mathcal R} \prod_{n_1 , n_2 \in \mathbb N}\left[  - \epsilon^{(-)}_1 \left(n_1 + \frac12 \right) - \epsilon^{(-)}_2 \left (n_2 + \frac12 \right) + i \, \rho(\Phi_0') \right] ~ ,  \\
  & = \prod_{\rho \in \mathcal R} \left[ \Gamma_2 ( i \, \rho(\Phi_0') - ( (\epsilon^{(-)}_1 + \epsilon^{(-)}_2)/2 )  | - \epsilon^{(-)}_1 , - \epsilon^{(-)}_2 ) \right]^{-1}  ~ .
\end{split}
\end{equation}
These  are  1-loop contribution to the hypermultiplet partition function at a minus fixed point.  For example, in the case of the squashed four-sphere studied in~\cite{Hama:2012bg}, the regularization chosen for the minus point contribution is\footnote{In the notation of~\cite{Hama:2012bg},  the regularization appears to be $--$, while in ours is $-+$, as reported in the main text. } $-+$:  
\begin{equation}
 \tilde Z^\text{HM}_{\epsilon^{(-)}_1 , \epsilon^{(-)}_2}( a_0' , k_- )  \equiv  \left[ Z^\text{HM}_{\epsilon^{(-)}_1 , \epsilon^{(-)}_2}( a_0' , k_- ) \right]_{-+} ~ . 
 \end{equation}
Using the results of~\cite{Festuccia:2018rew}, we can write down the complete partition function for an equivariantly twisted $\mathcal N=2$ gauge theory coupled to matter with $p$ (respectively $(l-p)$) plus (minus) fixed points, giving formula (\ref{eq: intromasterformula}).

\subsection{Example: hypermultiplets on squashed $S^4$ }
As an example, let us  apply (\ref{eq: intromasterformula}) to the specific case of hypermultiplets on a squashed four-sphere $S^4_{\ell , \tilde \ell}$ studied in~\cite{Hama:2012bg}. This manifold possesses a plus and a minus fixed point, respectively dubbed north and south pole. In fact, $S^4_{\ell , \tilde \ell}$ is a four-dimensional ellipsoid embedded in $\mathbb R^5$ according to
\begin{equation}
  \frac{x_1^2 + x_2^2}{\ell^2} + \frac{x_3^2 + x_4^2}{\widetilde\ell^2} + \frac{  x_5^2}{r^2} = 1 ~ , 
\end{equation}       
where $\ell, \widetilde \ell $ are the lengths of the ellipsoid axis, while $r$ is its radius. If $\ell = \widetilde \ell = r $, one recovers the round four-sphere $S^4$, whose isometry group is $SO(5)$. For arbitrary $\ell, \widetilde \ell $, the group $SO(5)$ is broken to $SO(2)\times SO(2)$, which is a real torus action rotating $(x_1,x_2)$ and $(x_3,x_4)$.  At the fixed points, the equivariant parameters $\epsilon_1^{(i)}$ and $\epsilon_2^{(i)}$ are related to the lengths of the ellipsoid axis as follows: 
\begin{center}
  \begin{tabular}{ c | c | c }
     & N & S   \\ \hline
    $\epsilon_1^{(i)}$ & $ \ell^{-1} $ & $ \ell^{-1} $   \\ \hline
    $\epsilon_2^{(i)}$ &  $ \tilde \ell^{-1} $ & $ - \tilde \ell^{-1}  $ 
  \end{tabular}
\end{center}
Indeed, we can make contact with the previous subsection by setting $z_1 = x_1 + i x_2 $ and $z_2 = x_3 + i x_4$, so that at the north pole of $S^4_{\ell , \tilde \ell}$ the torus action becomes a complex torus action $U(1) \times U(1)$ generated by a Killing vector (\ref{eq: locppkv}) with $\epsilon^{(+)}_1 = \ell^{-1}$ and $\epsilon^{(+)}_2 = \tilde \ell^{-1}$. On $S^4_{\ell , \tilde \ell}$ we have vanishing fluxes, then $k_+ = k_-=0$, and for a hypermultiplet in the representation $\mathcal R $ of the gauge group we find
\begin{equation}
  Z^\text{HM}_{S^4_{\ell , \tilde \ell} }( a_0  ) |_\text{north pole}= \left[  Z^\text{HM}_{\ell^{-1}  , \tilde \ell^{-1} }( a_0 , 0 ) \right]_{++} = \prod_{ \rho \in \mathcal R}\Gamma_2( i \, \rho(a_0) + \frac{\ell^{-1} + \tilde \ell^{-1}}2  | \ell^{-1} , \tilde \ell^{-1} ) ~ .
\end{equation}
Analogously, at the south pole we have a $U(1) \times U(1)$ generated by a Killing vector (\ref{eq: locmpkv}) with $\epsilon^{(-)}_1 =  \ell^{-1}$ and $\epsilon^{(-)}_2 = - \tilde \ell^{-1}$. Consequently,
\begin{equation}
  Z^\text{HM}_{S^4_{\ell , \tilde \ell} }( a_0  ) |_\text{south pole}=  \left[ Z^\text{HM}_{\ell^{-1}  , - \tilde \ell^{-1} }( a_0 , 0 ) \right]_{-+} = \prod_{ \rho \in \mathcal R}\Gamma_2(  i \, \rho(a_0) -   \frac{\ell^{-1} + \tilde \ell^{-1}}2 | - \ell^{-1} , -  \tilde \ell^{-1} ) ~ .
\end{equation}
Combining the two contributions we obtain a $\Upsilon$-function depending on the Coulomb branch parameter $a_0$ and on the lengths of the ellipsoid axis $\ell, \tilde \ell$:
\begin{align}
  Z^\text{HM}_{S^4_{\ell , \tilde \ell} }( a_0  )    & =     Z^\text{HM}_{S^4_{\ell , \tilde \ell} }( a_0  ) |_\text{north pole}  \times    Z^\text{HM}_{S^4_{\ell , \tilde \ell} }( a_0  ) |_\text{south pole}   ~ , \nonumber \\
  & =  \prod_{\rho \in \mathcal R} \Upsilon_\beta\left( i \sqrt{ \ell \tilde \ell} \,  \rho( a_0  ) +  \frac 12(\beta + \beta^{-1})  \right)^{-1} ~ ,  & \beta = \sqrt{\ell/\tilde \ell} ~ , 
\end{align}
which matches~\cite{Hama:2012bg}. We recall that the definition of the  $\Upsilon$-function is
 \begin{align}
  \Upsilon_\beta ( x ) = \prod_{m,n \in \mathbb N} \left( m \beta + n \beta^{-1} + x \right) \left( m \beta + n \beta^{-1} + \beta + \beta^{-1} - x \right)   ~ .
 \end{align}

\subsection{Example: hypermultiplets on  $S^2 \times S^2$ }

As another example, we apply (\ref{eq: intromasterformula}) to hypermultiplets  defined on the product of two spheres of radii $\epsilon^{-1}_1 , \epsilon^{-1}_2$, which we denote by $S^2_{\epsilon_1} \times S^2_{\epsilon_2} $. Gauge theories on  manifolds with such a topology were studied e.g. in~\cite{Bawane:2014uka}, as well as in~\cite{Festuccia:2016gul} by dimensional reduction from five-dimensional  toric Sasaki–Einstein manifolds. There are four fixed points on the  manifold $S^2_{\epsilon_1} \times S^2_{\epsilon_2} $,    corresponding to the four combinations NN, NS, SN, SS of north (N) and south (S) poles of the two spheres. Here, we consider  NN and NS being plus fixed points, while SN and SS being minus fixed points. At the fixed points, the local equivariant parameters $\epsilon_1^{(i)}$ and $\epsilon_2^{(i)}$ are related to $\epsilon_1$ and $\epsilon_2$ as follows:
\begin{center}
  \begin{tabular}{ c | c | c | c | c }
     & NN & NS & SN & SS  \\ \hline
    $\epsilon_1^{(i)}$ & $\epsilon_1$ & $\epsilon_1$ & $- \epsilon_1$ & $ - \epsilon_1$  \\ \hline
    $\epsilon_2^{(i)}$ &  $\epsilon_2 $ & $ - \epsilon_2 $ & $ \epsilon_2 $ & $ - \epsilon_2$
  \end{tabular}
\end{center}
Using a regularization consistent with~\cite{Festuccia:2016gul}, the  hypermultiplet contribution at  plus fixed points  is
\begin{align}
  Z^\text{HM}_{ S^2_{\epsilon_1} \times S^2_{\epsilon_2}  }( \Phi_0  ) |_\text{NN, NS}
  & = \left[  Z^\text{HM}_{ \epsilon_1  ,  \epsilon_2 }( \Phi_0 |_\text{NN} ) \right]_{++} \times  \left[  Z^\text{HM}_{ \epsilon_1  , - \epsilon_2 }( \Phi_0 |_\text{NS} ) \right]_{+-} ~ ,  \nonumber\\
  &  = \prod_{\rho \in \mathcal R} \frac{\Gamma_2 ( i \, \rho(\Phi_0 |_\text{NN})  +  \frac{\epsilon_1 + \epsilon_2}2  |  \epsilon_1 , \epsilon_2 )   }{ \Gamma_2 ( i \, \rho(\Phi_0 |_\text{SS} ) + \frac{\epsilon_1 + \epsilon_2}2   |   \epsilon_1 ,  \epsilon_2 ) }  ~ ,
\end{align}
while the  hypermultiplet contribution at  minus fixed points reads  
\begin{align}
  \tilde Z^\text{HM}_{ S^2_{\epsilon_1} \times S^2_{\epsilon_2}  }( \Phi_0'  ) |_\text{SN, SS}
  & =  \left[  Z^\text{HM}_{ - \epsilon_1  ,   \epsilon_2 }( \Phi_0' |_\text{SN} ) \right]_{++}  \times \left[  Z^\text{HM}_{ - \epsilon_1  ,  - \epsilon_2 }( \Phi_0' |_\text{SS} ) \right]_{+-} ~ ,  \nonumber \\
  & = \prod_{\rho \in \mathcal R} \frac{ \Gamma_2 (  i \, \rho(\Phi_0' |_\text{SS}) + \frac{ - \epsilon_1 + \epsilon_2}2 )   | -  \epsilon_1 ,   \epsilon_2 ) }{\Gamma_2 (  i \, \rho(\Phi_0' |_\text{SN}) + \frac{ - \epsilon_1 + \epsilon_2}2   |  - \epsilon_1 , \epsilon_2 )} ~ .
\end{align}
 In absence of fluxes, $\Phi_0 = \Phi_0'= a_0$ at any fixed point and the   partition function trivializes:
 \begin{align}
  & Z^\text{HM}_{ S^2_{\epsilon_1} \times S^2_{\epsilon_2}  }( a_0 ) = 1 ~ .
 \end{align}

\section*{Acknowledgments}
We thank Jian Qiu and Maxim Zabzine for illuminating discussions and for a critical  reading of the draft.  We also thank Luca Cassia for clarifying some issues relating to generalized ${\rm Spin}$ structures. G.F. acknowledges the support of the ERC STG Grant 639220 and of Vetenskapsr\aa{}det under grant 2018-05572. A.G. receives support partly by Vetenskapsr\aa{}det under grant \#2016-03503 and by the Knut and Alice Wallenberg Foundation under grant Dnr KAW 2015.0083.  The work of K.P. is supported by the grant ``Geometry and Physics'' from the Knut and Alice Wallenberg foundation. A.P. and L.R. are supported by the ERC STG Grant 639220.

\appendix

\section{Notation and Conventions}\label{app:conv}
\label{sec:conventions}
Here we collect the relevant formulas used in the main text and a summary of our conventions. These are based on those of~\cite{Wess:1992cp}, adapted to Euclidean signature.

\subsection{Flat Euclidean space and Dirac spinors}
We define the Levi--Civita symbol as ~$\epsilon_{1234} = 1$. The rotation group is~$SO(4) \sim {\rm Spin}(4) = SU(2)_+ \times SU(2)_-$.  Left-handed spinors are $SU(2)_+$ doublets and are denoted by undotted indices~$\zeta_\alpha$. Right-handed spinors $\bar \zeta_{\dot\alpha}$ are doublets under~$SU(2)_-$ and carry a bar as well as dotted indices. In Euclidean signature, $SU(2)_+$ and~$SU(2)_-$ are not related by complex conjugation, hence $\zeta$ and~$\bar \zeta$ are independent spinors. We raise and lower dotted and undotted indices by acting on the left with the tensors $\epsilon_{\alpha \beta} $ and $\epsilon_{\dot \alpha \dot \beta} $ , where $\epsilon^{12} = \epsilon_{21} = \epsilon^{\dot 1 \dot 2} = \epsilon_{\dot 2 \dot 1} = + 1$. For instance, $\zeta^\alpha=\epsilon^{\alpha \beta}\zeta_\beta$ and $\bar \zeta^{\dot \alpha}=\epsilon^{{\dot \alpha} {\dot \beta}}\bar \zeta_{\dot \beta} $.  The~$SU(2)_+$ invariant inner product of~$\zeta$ and~$\eta$ is~$\zeta \eta=\zeta^\alpha \eta_\alpha$. The~$SU(2)_-$ invariant inner product of~$\bar \zeta$ and~$\bar \eta$ is given by~${\bar \zeta} \bar \eta={\bar \zeta}_{{\dot \alpha}} \bar \eta^{\dot \alpha}$.  We introduce the sigma matrices
\begin{equation}
  \label{sigmamat}
  \sigma^\mu_{\alpha \dot{\alpha}} = (\vec{\sigma}, -i \id )~,
  \qquad
  \bar \sigma^{\mu\dot{\alpha} \alpha} = (-\vec{\sigma}, -i \id )~, 
\end{equation}
where ~$\vec{\sigma} = (\sigma^1, \sigma^2, \sigma^3)$ is a vector whose components are Pauli matrices. Four-dimensional sigma matrices satisfy the reality conditions $(\sigma_\mu)^\dagger = - \bar \sigma_\mu$. Furthermore,
\begin{equation}
  \sigma_\mu\bar \sigma_\nu + \sigma_\nu \bar \sigma_\mu = -2\delta_{\mu\nu}~, \qquad
  \bar \sigma_\mu \sigma_\nu + \bar \sigma_\nu \sigma_\mu = -2\delta_{\mu\nu}~.
\end{equation}
We also define the matrices
\begin{equation}
  \sigma_{\mu\nu} = \frac{1}{4} (\sigma_\mu \bar \sigma_\nu - \sigma_\nu\bar \sigma_\mu)~, \qquad
  \bar \sigma_{\mu\nu} = \frac{1}{4} (\bar \sigma_\mu \sigma_\nu - \bar \sigma_\nu \sigma_\mu)~.
\end{equation}
The latter fulfill self-duality (or anti self-duality) properties:
\begin{equation}
  \frac{1}{2} \epsilon_{\mu\nu\rho\lambda} \sigma^{\rho \lambda } = \sigma_{\mu\nu}~, \qquad
  \frac{1}{2} \epsilon_{\mu\nu\rho\lambda} \bar{\sigma}^{\rho\lambda} = - \bar \sigma_{\mu\nu}~.
\end{equation}
We can use these matrices to separate the $(2,0)$ and $(0,2)$ components of a two-form $\omega$ as follows:
\begin{equation}
  \omega^{+}_{\alpha \beta} = \frac{1}{2} \omega_{\mu\nu}\sigma^{\mu\nu}_{\alpha \beta}~,\qquad
  \omega^{-}_{{\dot \alpha} {\dot \beta}}= \frac{1}{2} \omega_{\mu\nu}\bar \sigma^{\mu\nu}_{{\dot \alpha} {\dot \beta}}~.
\end{equation}

A Dirac spinor $\Psi$ contains a left-handed Weyl spinor $\psi_\alpha$ and a right-handed one $\tilde \psi^{\dot \alpha}$. They are arranged as
\begin{equation}
  \Psi=\begin{pmatrix}
    \psi_\alpha  \\
    \tilde\psi^{\dot \alpha}
  \end{pmatrix}~. 
\end{equation}
Correspondingly, the adjoint spinor $\overline \Psi$ is 
\begin{equation}
  \overline \Psi= \begin{pmatrix}
    \psi^\alpha && \!\!\!\!\!\!\tilde \psi_{\dot \alpha} 
  \end{pmatrix} = (C\Psi)^T = - \Psi^T C ~ ,
\end{equation}
where $T$ denotes transposition and $C$ is the (skew-symmetric) charge-conjugation matrix
\begin{align}
  & C = \begin{pmatrix}
    \epsilon^{\alpha  \beta}  && 0 \\
    0 &&  \epsilon_{\dot \alpha \dot \beta}
  \end{pmatrix} ~ ,     &   C^{-1} = \begin{pmatrix}
    \epsilon_{\alpha  \beta}  && 0 \\
    0 &&  \epsilon^{\dot \alpha \dot \beta}
  \end{pmatrix} ~ .
\end{align}
In general, the adjoint $\overline \Psi$ of a spinor $\Psi$, is not related to the conjugate spinor $\Psi^*$.  The ${\rm Spin}(4)$-invariant product between two Grassmann-odd Dirac spinors $\Psi_1, \Psi_2$ reads
\begin{equation}
  \overline \Psi_1\Psi_2 = - \Psi_1^T C \Psi_2   = \psi_1^\alpha \psi_{2\alpha}
  + \tilde \psi_{1 \dot \alpha} \tilde \psi_2^{\dot \alpha}~.
\end{equation}
The Clifford algebra is generated by Dirac matrices $\gamma_\mu$ in chiral representation, namely
\begin{align}
  & \gamma_\mu = \begin{pmatrix}
    0 &&   \sigma_\mu  \\
    \bar \sigma_\mu   &&   0
  \end{pmatrix} ~ , \qquad\{ \gamma_\mu , \gamma_\nu\} = - 2 g_{\mu\nu} ~ , & C\gamma_\mu C^{-1} = - (\gamma_\mu)^T  ~ ,
\end{align}
where $g_{\mu\nu}$ is the spacetime metric. The chirality matrix $\gamma_5$ and the ${\rm Spin}(4)$ generators $\gamma_{\mu\nu}$ are
\begin{align}
  & \gamma_5 = -  \gamma_1 \gamma_2 \gamma_3 \gamma_4 = \begin{pmatrix}
    \id  &&  0  \\
    0 && -  \id 
  \end{pmatrix}~,  
         &   \gamma_{\mu\nu} =\frac14 ( \gamma_{\mu}\gamma_{\nu} - \gamma_{\nu}\gamma_{\mu} ) =  \begin{pmatrix}
           \sigma_{\mu\nu}  && 0 \\
           0 &&  \bar \sigma_{\mu\nu}
         \end{pmatrix} ~ .
\end{align}
In particular, $C \gamma_5 C^{-1} = (\gamma_5)^T$, while $\sigma_{\mu\nu}$ and $\bar \sigma_{\mu\nu}$ generate $SU(2)_+$ and $SU(2)_-$ respectively.  The matrix $\gamma_5$ is used to construct the standard chiral projectors
\begin{align}
  & L = \frac12 \left ( \id + \gamma_5 \right) ~ ,
  & R = \frac12 \left ( \id - \gamma_5 \right) ~ .
\end{align}
Bilinears of Grassmann-odd  spinors  satisfy
\begin{align}
  \overline \Psi_1 \Psi_2 =   \overline \Psi_2 \Psi_1 ~ , \qquad
  \overline \Psi_1 \gamma_\mu \Psi_2 = - \overline \Psi_2 \gamma_\mu \Psi_1 ~ , \qquad
  \overline \Psi_1 \gamma_5 \Psi_2 =    \overline \Psi_2 \gamma_5 \Psi_1    ~ .
\end{align}
Bilinears of Grassmann-even spinors fulfill the same identities with an additional minus sign on the right-hand side.

\subsection{Differential geometry}
Greek letters~$\mu, \nu, \ldots$ are used to denote curved indices and Latin letters~$a, b, \ldots$ to denote frame indices. Let~${e^a}_\mu$ be the orthonormal vielbein corresponding to the metric~$g_{\mu\nu}$. We denote the Levi--Civita connection by~$\nabla_\mu$. The corresponding spin connection is
\begin{equation}
  {\omega_{\mu a}}^b = {e^b}_\nu \nabla_\mu {e_a}^\nu~.
\end{equation}
The Riemann tensor is
\begin{equation}
  {R_{\mu\nu a}}^b = \partial_\mu {\omega_{\nu a}}^b - \partial_\nu {\omega_{\mu a}}^b
  + {\omega_{\nu a}}^c {\omega_{\mu c}}^b - {\omega_{\mu a}}^c {\omega_{\nu c}}^b~.
\end{equation}
The Ricci tensor is~$R_{\mu\nu} = {R_{\mu\rho\nu}}^\rho$, and~$R = {R_\mu}^\mu$ is the Ricci scalar. In these conventions a round sphere has negative Ricci scalar.

The covariant derivative acts on spinors~$\zeta$ and~$\bar \zeta$ as
\begin{equation}
  \nabla_\mu \zeta = \partial_\mu \zeta + \frac{1}{2}\omega_{\mu a b } \sigma^{ab} \zeta~, \qquad
  \nabla_\mu \bar \zeta = \partial_\mu \bar \zeta + \frac{1}{2} \omega_{\mu a b } {\bar \sigma}^{ab} \bar \zeta~.
\end{equation}

The Lie derivative of spinors along a Killing vector $v$ is given by~\cite{Kosmann1971}
\begin{equation}
  \mathcal{L}_v \zeta = v^\mu \nabla_\mu \zeta
  - \frac{1}{2} (\partial_\mu v_\nu) \sigma^{\mu\nu} \zeta~,\qquad
  \mathcal{L}_v \bar \zeta = v^\mu \nabla_\mu \bar \zeta
  - \frac{1}{2} (\partial_\mu v_\nu) \bar \sigma^{\mu\nu} \bar \zeta~.
\end{equation}

\section{Supergravity background solutions}
\label{sec:sugra-background-solutions}
In this appendix, we write down explicitly the supergravity background on which the Killing spinor equations are solved by the Killing spinor constructed in section~\ref{sec:constr-killing-spinors}. This is a summary of the results of~\cite{Festuccia:2018rew}.

The background supergravity fields are the metric $g_{\mu\nu}$, a $SU(2)_R$ connection ${{V_{\mu}}^i}_j$, a one form $G_\mu$, a two-form $W_{\mu\nu}$, a scalar $N$, a closed two-form $F_{\mu\nu}$, and a scalar $S_{ij}$ transforming as an $SU(2)_R$ triplet. The Killing spinors do not determine the supergravity background completely. In the formulas below, this freedom is parametrized by two one-forms. One is $G_\mu$ and is arbitrary, the other is denoted $b_\mu$ and satisfies $\imath_{v}b=0$. The reality conditions for the Killing spinors~\eqref{eq:reality-killing} are compatible with the following behavior of the background supergravity fields under complex conjugation:
\begin{equation}\label{eq:sugra-reality}
  ({{V_\mu}^i}_j)^*={{V_\mu}^j}_i~,\quad
  N^*=N~,\quad
  G_\mu^*=-G_\mu~,\quad
  W_{\mu\nu}^*=-W_{\mu\nu}~,\quad
  {\cal F}_{\mu\nu}^*=-{\cal F}_{\mu\nu}~,\quad
  S_{i j}^*=S^{i j}~.
\end{equation}

We will make use of various spinor bilinears. Besides the $SU(2)_R$ singlets $s,\tilde s $ and $v$ defined in~\eqref{stsdef} and~\eqref{eq:Killing-vector} and used extensively throughout the paper, we introduce
\begin{equation}
  v^{(ij)}_\mu= \zeta^i \sigma_\mu \bar\chi^j+\zeta^j \sigma_\mu \bar\chi^i~,\qquad
  \Theta_{\mu\nu}^{(ij)}=\zeta^i \sigma_{\mu\nu}\zeta^j~,\qquad
  \widetilde \Theta^{(ij)}_{\mu\nu}=\bar\chi^i \bar \sigma^{\mu\nu} \bar \chi^j~.
\end{equation}
We also define the combination
\begin{equation}
  \label{thetb}
  \widehat\Theta^{i j}_{\mu\nu}= 2{s+\tilde s \over s^2+\tilde s^2}
  \left(\Theta^{i j}_{\mu\nu}+\widetilde\Theta^{i j}_{\mu\nu}\right)~.
\end{equation}
The two-form $W_{\mu\nu}$ and the $SU(2)_{R}$ connection $(V_\mu)_{i j}$ are given by:
\begin{align}
  W_{\mu\nu} =& {i \over s+\tilde s} (\partial_\mu v_\nu-\partial_\nu v_\mu)
                -{2i \over (s+\tilde s)^2} {\epsilon_{\mu\nu\rho}}^{\lambda}v^\rho \partial_\lambda(s-\tilde s)
                -{4 \over s+\tilde s}{\epsilon_{\mu\nu\rho}}^\lambda v^\rho G_\lambda \nonumber\\
              & +{s-\tilde s\over (s+\tilde s)^2}{\epsilon_{\mu\nu\rho}}^\lambda v^\rho b_\lambda
                +{1\over s+\tilde s}(v_{\mu} b_{\nu} -v_{\nu}b_{\mu})~, \\
  (V_\mu)_{i j} = &{4\over s+\tilde s}\left(\zeta_{(i}\nabla_\mu \zeta_{j)}
                    +\bar \chi_{(i}\nabla_\mu \bar \chi_{j)}\right)
                    +{4\over s+\tilde s} \left(2 i G_\nu-{\partial_\nu(s-\tilde s)\over (s+\tilde s)}\right)
                    {(\Theta_{i j}-\widetilde \Theta_{i j})^\nu}_\mu\nonumber\\
              &+ {4i\over (s+\tilde s)^2}b_\nu ( \tilde s\, \Theta_{i j}+s\,\widetilde  \Theta_{i j} )^\nu{}_\mu~.
\end{align}
The graviphoton field strength $\mathcal{F}_{\mu\nu}$ is:
\begin{equation}
  {\mathcal F}_{\mu\nu} =
  i \partial_{\mu}\Big({s+\tilde s-K\over s \tilde s} v_\nu\Big)
  -i\partial_{\nu}\Big({s+\tilde s-K\over s \tilde s} v_\mu\Big)~.
\end{equation}
Here $K$ is the constant defined in section~\ref{secspin} and such that
$(s+\tilde s-K)/(s^2 \tilde s^2)$ is smooth at the fixed points of $v$.  The
scalars $S_{ij}$ are:
\begin{align}
  S_{i j}=4 i  {s^2 +\tilde{s}^2 \over (s+\tilde s)^3}
  & \widehat\Theta_{i j}^{\mu\nu}\partial_\mu v_\nu
    - 2i{ s+\tilde s-K\over (s\, \tilde s)^2}
    (\tilde s\,\Theta_{i j}+s\, \widetilde \Theta_{ij})^{\mu\nu}\partial_\mu v_\nu \nonumber\\
  &\quad \,\,+{2\over s+\tilde s} \Big({4}G_\mu-{s-\tilde s\over s+\tilde s} b_\mu
    -{i\over 2} {s-\tilde s\over s \tilde s}\partial_\mu(s+\tilde s)\Big) v^{\mu}_{i j}~.
\end{align}
Finally, the combination $R/6 - N$ (where $R$ is the Ricci scalar) is given by:
\begin{align}
  \Big({R\over 6}-N\Big)=
  &~{s-\tilde s\over(s+\tilde s )^2 }\Box(s-\tilde s)
    +{ \partial_{[\mu} v_{\nu]}\partial^{[\mu}v^{\nu]}\over (s+\tilde s)^2}
    -{1\over 2}{s-\tilde s \over (s+\tilde s)^3} \epsilon^{\mu\nu\rho\lambda}(\partial_\mu v_\nu)(\partial_\rho v_{\lambda})+\nonumber\\
  &-{4 \epsilon^{\mu\nu\rho\lambda}\over (s+\tilde s)^3} v_\mu(\partial_\nu v_\rho)\partial_\lambda(s-\tilde s)
    +{2 s \tilde s\over (s+\tilde s)^4}\partial_\mu(s-\tilde s)\partial^\mu (s-\tilde s)+\nonumber\\
  &-2{s -\tilde s\over (s+\tilde s)^3}\partial_\mu(s-\tilde s)\partial^\mu (s+\tilde s)
    -{2i s \tilde s\over (s+\tilde s)^2} \nabla^\mu b_\mu+{2(s \tilde s)^2\over (s+\tilde s)^4} b_\mu b^\mu+\nonumber\\
  &-i{s-\tilde s\over (s+\tilde s)^3} \epsilon^{\mu\nu\rho\lambda}(\partial_\mu v_\nu)v_\rho b_\lambda
    +3i{s-\tilde s\over (s+\tilde s)^3} b^\mu (\tilde s \partial_\mu s-s\partial_\mu \tilde s)+\nonumber\\
  &+2i{ s^2+ \tilde s^2 \over (s+\tilde s)^4}b^\mu \partial_\mu(s \tilde s)
    -2i{s-\tilde s \over s+\tilde s}\nabla^\mu G_\mu
    +4{s^2+{\tilde s}^2\over (s+\tilde s )^2} G^\mu G_\mu
    +{8 (v^\mu G_\mu)^2\over (s+\tilde s)^2}+\nonumber\\
  &+4{s\tilde s} {s-\tilde s \over(s+\tilde s)^3}G_\mu b^\mu
    +{ 4i\epsilon^{\mu\nu\rho\lambda} \over(s+\tilde s)^2} (\partial_\mu v_\nu)v_\rho G_\lambda
    +4i{s-\tilde s \over (s+\tilde s)^3} G^\mu\partial_\mu(s \tilde s)~.
\end{align}

\section{Cohomological variables for vector multiplet}\label{app:twisvec}
The component fields in the $\mathcal{N}=2$ vector multiplet are the complex scalar $X$, a gauge field $A_{\mu}$, two gauginos $\lambda_{i\alpha}$ and $\tilde{\lambda}_{\dot{\alpha}}^{i}$ and an auxiliary scalar field $D_{ij}$.  These can be recast in the following twisted fields (see~\cite{Festuccia:2018rew}):
\begin{equation}
  \begin{aligned}
    \label{twtd}
    &\eta =\zeta_i \lambda^i+\bar \chi^i \bar \lambda_i~,\\
    &\varphi = -i(X-\bar X)~,\\
    &\Psi_\mu = \zeta_i \sigma_\mu \bar \lambda^i+\bar \chi^i\bar \sigma_\mu \lambda_i~,\\
    &\phi = \tilde s X+ s \bar X~,\\
    &\chi_{\mu\nu} = 2{s+\tilde s\over s^2+\tilde s^2}\left[{ \bar \chi^i\bar\sigma_{\mu\nu}\bar \lambda_i}
      -{ \zeta_i \sigma_{\mu\nu} \lambda^i}
      +{1\over s + \tilde s} (v_{\mu} \Psi_\nu-v_\nu \Psi_\mu)\right]~,\\
    &H_{\mu\nu} = {(P_+)}_{\mu\nu}^{\rho \lambda}\left[  \hat \Theta^{i j}_{\rho\lambda} D_{i j}
      - F_{\rho\lambda}+i{X+\bar X\over s+\tilde s} (\partial_\rho v_\lambda-\partial_\lambda v_\rho)+ \right. \\
    &\qquad \left.-{2i\over s+\tilde s} {\epsilon_{\rho\lambda\gamma}}^{\delta}v^\gamma\left(\Big(D_\delta
        -2iG_\delta -i{\tilde s\over s+\tilde s}\, b_\delta \Big)X
        -\Big(D_\delta+2iG_\delta -i {s\over s+\tilde s}\, b_\delta\Big)\bar X\right) \right]~.
  \end{aligned}
\end{equation}
Here we used $ \hat \Theta^{i j}_{\rho\lambda} $ defined in~\eqref{thetb} and the projector $P_+$ defined in~\eqref{defpp} .

All the fields~\eqref{twtd} are differential forms in the adjoint of the gauge group. They are all singlets under the $SU(2)_R$ symmetry. It follows from the definitions above that the two-forms $\chi_{\mu\nu}$ and $H_{\mu\nu}$ are in the image of the projector $P_+$~.

The change of variables~\eqref{twtd} has a smooth inverse,
\begin{equation}
  \label{intwtd}
  \begin{aligned}
    &X={1 \over s+\tilde s}\,(\phi +i\, s \, \varphi) ~,\qquad
    \bar X={1 \over s+\tilde s}\,(\phi - i\, \tilde s \,\varphi)~,\\
    & \bar \lambda_i ={1\over s+\tilde s}\, \left[  {s^2 +\tilde{s}^2 \over s+\tilde s}\bar \chi^j\,
      \widehat\Theta_{i j}^{\mu\nu}
      \chi_{\mu\nu}
      +\bar \sigma^\mu \zeta_i\Psi_\mu+{\bar \chi_i \eta}\right]~,\\
    &\lambda_i ={1\over s+\tilde s} \,\left[  {s^2 +\tilde{s}^2 \over s+\tilde s}\zeta^j\,
      \widehat\Theta_{i j}^{\mu\nu}
      \chi_{\mu\nu}
      -\sigma^\mu \bar \chi_i\Psi_\mu-{\zeta_i \eta}\right]~,\\
    &D_{i j}=4{s^2+\tilde s^2\over (s+\tilde s)^2} \hat \Theta^{\mu\nu}_{i j}(H_{\mu\nu}-\ldots)~,
  \end{aligned}
\end{equation}
where in the last formula the ellipsis stand for terms in $H_{\mu\nu}$ that are not proportional to $D_{ij}$ (see~\eqref{twtd}).

\bibliographystyle{jhep}
\bibliography{hyper_twist}

\end{document}